# DETERMINATION OF DISTANCE FROM TIME DILATION OF COSMOLOGICAL GAMMA-RAY BURSTS

E. E. Fenimore and J. S. Bloom

Los Alamos National Laboratory, Los Alamos, New Mexico 87545, USA

E-mail: efenimore@lanl.gov

## ABSTRACT

The isotropic distribution of gamma-ray bursts as observed with the Burst and Transient Experiment (BATSE) strongly suggests that the bursts are at cosmological distances. At such distances, the expansion of the universe should redshift the spectra and stretch the temporal structure. Indeed, such time dilation has been observed through a variety of analyses and over all observed time scales in gamma-ray bursts. We relate the observed peak intensities, spectral shapes, and time dilation to absolute distance. We include the uncertainties in our knowledge of the intrinsic spectrum and correct for the coupling between the spectral shape and the temporal structure. Assuming a $q_0 = 1/2$ cosmology, the reported time dilation between the dimmest BATSE bursts and the bright BATSE bursts (a factor of $\sim 2$) requires a standard candle luminosity of $\sim 10^{52}$ erg s$^{-1}$, which translates into a redshift of $> 6$ for the BATSE dimmest bursts rather than a redshift of 1 or 2 as previously reported. An alternative method to determine the distance to cosmological GRBs is to use the log $N$-log $P$ distribution. The large luminosity and distance determined from the time dilation is inconsistent with the observed log $N$-log $P$ distribution (which requires a luminosity of $4.6 \times 10^{50}$ erg s$^{-1}$ and $z \sim 0.8$) unless there is very strong evolution. Cosmologies with $q_0 \neq 1/2$ give similar results. The implied distance for the dimmest bursts is beyond where galaxies are thought to form. If true, the gamma-ray bursts would be orphans: no known objects would have the same distance scale. We conclude that either a large fraction (65%) of the observed time dilation between the BATSE bright and dimmest bursts is intrinsic to the bursts or that there are strong evolutionary effects in the log $N$-log $P$ distribution and that it is only a coincidence that log $N$-log $P$ shows a $-3/2$ power law at high intensities.

*Subject headings:* gamma rays: bursts – cosmology: theory

# 1. INTRODUCTION

Until 1991 there was a strong consensus that gamma-ray bursts (GRBs) originated from within our Galaxy. More recently, the Burst and Transient Experiment (BATSE) on the *Compton Gamma-Ray Observatory* discovered that there is a dearth of weak events despite the fact that the bursts are isotropic (Meegan et al. 1992). This is strong evidence that the bursts originate far from the influence of the mass distribution of our Galaxy and even the distribution of nearby clusters of galaxies. This has resulted in a great debate as to whether gamma-ray bursts are within an extended halo of our Galaxy or at cosmological distances. Most observations are either inconclusive or controversial. The isotropic distribution could be formed by a halo. Recently, halos with properties consistent with the BATSE observations have been modelled using improved neutron star dynamics (Podsiadlowski, Rees, and Ruderman 1994). Harmonically spaced cyclotron lines (Murakami et al. 1988, Fenimore et al. 1988) might be a signature of galactic neutron stars but it is unclear how such lines can be formed in the super relativistic flows required for halo neutron stars (although, see Miller et al. 1991). Even though no lines are detected with BATSE, its sensitivity to such lines is not sufficient to rule out their existence (Band et al. 1994). A log $N$-log $P$ distribution gives the number of events ($N$) with a peak intensity $P$ (photons s$^{-1}$ cm$^{-2}$) above some value and is sensitive to the distribution of events in space. The log $N$-log $P$ distribution expected from a cosmological distribution has been successfully fit to the combined BATSE - Pioneer Venus Orbiter (PVO) observations (Fenimore et al. 1993). However, this is not conclusive since galactic distributions could also fit the data.

Dimmer cosmological bursts should not only be redshifted compared to brighter (presumably nearer) bursts, but the expansion of the universe also stretches the bursts in time. Recently, time dilation has been discovered in BATSE GRBs (Norris 1994, Norris et al. 1994, Davis et al. 1994). The BATSE bursts were assigned to a brightness class depending on their peak count rate: the "bright," "dim," and "dimmest" class correspond to 18000 to 25000, 2400 to 4500, and 1400 to 2400 cts s$^{-1}$ in BATSE, respectively. (Events with an intermediate count rate are not used since the time dilation effects are largest for well-separated classes.) There have been a variety of tests that claimed a measured time dilation between bright and dimmest events. The total-count test, wavelet-power test, and aligned-peak test find a factor of 2 dilation between the dimmest and bright BATSE events (Norris 1994, Norris et al. 1994). However, Mitrofanov et al. (1994) finds no time dilation in BATSE using the aligned-peak test (Mitrofanov et al. 1993) and Band (1994) has criticized these tests. Davis et al. (1994) found that a test based on the average pulse width required a dilation factor of 1.8, a test based on the number of counts within individual peaks required a factor of 2.2, and a test based on the total number of counts in the bursts required a factor of 2.2. Davis et al. 1994 concluded that a time dilation of 2.0 best explained the data. Fenimore et al. (1995b), using the average auto-correlation function, confirmed the time dilation of a factor of 2 between the bright and dim BATSE events discovered by Norris et al. (1994) and investigated and eliminated possible sources of systematic errors. Recently,

Norris et al. (1995) investigated GRB durations and found that the dimmest events were dilated by a factor of $2.2 \pm 0.2$. The dimmest BATSE bursts are also softer than the bright bursts as expected for redshifted cosmological bursts (Nemiroff et al. 1994).

The beauty of the time-dilation tests is that they can provide a measure of the distance scale to cosmological GRBs that is independent of the log $N$-log $P$ distribution yet must be consistent with it. The time-dilation tests typically use averages from the count rate time history of the bursts as a function of brightness, whereas the log $N$-log $P$ test uses the number of detected events. If these two rather independent tests are consistent, it would be strong evidence in favor of the cosmological explanation for the origin of GRBs. Indeed, Norris 1994, Norris et al. 1994, and Davis et al. 1994 claimed that the observed time dilation of $\sim 2$ implied that the dimmest bursts were at redshifts, $z$, of $\sim 1$, consistent with the redshift of the dimmest events found from the log $N$-log $P$ studies (Fenimore et al. 1993, Wickramasinghe et al. 1993).

In this paper, we find the relationship between time dilation and distance correcting for three effects. First, one must assume a spectral shape to convert intensity into a distance. In the past, simple spectral shapes (e.g., thermal bremsstrahlung or power law) were used *and* it was assumed that one knew the spectral shape at $z = 0$. In this paper, we will use an average over the observed shapes (cf. Band et al. 1993) coupled with their observed intensities to determine the distance (redshift) to the events that define the spectral shape. Second, the bright class is not necessarily at $z = 0$ (nor at the redshift of the events that defined the spectral shape) so that the time dilation depends on $(1+z_{\text{dimmest}})/(1+z_{\text{bright}})$ rather than $1+z_{\text{dimmest}}$ (Mitrofanov et al. 1994). And, third, since time histories are narrower at high energy for a given $z$, there is a correction based on the redshift of the spectrum.

## 2. RELATING INTRINSIC LUMINOSITY TO DISTANCE

As a first step in determining the redshift of a gamma-ray burst from observed time dilation, we develop a relationship between standard-candle luminosity, $L_0$, and redshift, $z$, for different brightness classes. The luminosity of a source in the detector bandpass depends strongly on the spectrum and, since observed spectral shape depends on the distance to the object, the intrinsic spectrum of a GRB object must be found. Rather than assume a specific spectral shape at $z = 0$ such as thermal bremsstrahlung, we take as our baseline spectra averages over the GRB spectra fit by Band et al. (1993). Each such burst has associated with it an observed intensity, $P_{i,B}$, and an observed spectral shape, $\phi_i(E)$, determined by three best-fit parameters provided by Band et al. (1993): power law indexes for low and high energies and a joining energy. The intensities, $P_{i,B}$, in units of photons cm$^{-2}$ s$^{-1}$, are given in the BATSE 2B catalog (Meegan et al. 1994) for the range 50 to 300 keV and are available for 50 of the Band et al. (1993) bursts.

The observed spectral shape, $\phi_i(E)$, will not necessarily come from a burst at $z \sim 0$, especially if $L_0$ is large. Therefore, for a given $L_0$, $P_{i,B}$, and $\phi_i(E)$, we first solve for the redshifts, $z_{i,B}$, of the events that are associated with the spectral shape. Let $\phi_i(E)$ be an observed spectral shape for a fixed standard candle luminosity, $L_0$. The normalization of the spectrum, $\phi_{0,i}$, is found from:

$$L_0 = \int_{30}^{2000} E\phi_{0,i}\phi_i\left(\frac{E}{1+z_{i,B}}\right)\frac{dE}{1+z_{i,B}}. \quad (1)$$

The energy range used in calculating the intrinsic luminosity of the source is taken as 30 to 2000 keV since we later compare the luminosity to that found by Fenimore et al. (1993) from log $N$-log $P$ studies. The $z = 0$ spectral shape is $\phi\left(\frac{E}{1+z_{i,B}}\right)$, whereas the observed spectrum is $\phi(E)$. The inclusion of the $1 + z_{i,B}$ factor corresponds to the blueshift of the $i\underline{\text{th}}$ baseline burst, that is, a burst that we use to define the intrinsic spectral shape.

The observed intensity of the $i\underline{\text{th}}$ baseline burst is given by:

$$P_{i,B} = \frac{L_0}{\int_{30}^{2000} E\phi_i\left(\frac{E}{1+z_{i,B}}\right)\frac{dE}{1+z_{i,B}}} \frac{\int_{50}^{300}\phi_i(E)dE}{4\pi R_{i,z}^2} \quad (2)$$

where the comoving distance is

$$R_{i,z} = \frac{c}{(1+z_{i,B})q_0^2 H_0}\left[q_0 z_{i,B} + (1-q_0)\left(1-(1+2z_{i,B}q_0)^{1/2}\right)\right], \quad (3)$$

$c$ is the speed of light, and $H_0$ is the Hubble constant which we take to be 75 km s$^{-1}$ Mpc$^{-1}$. Except for in section 6, we assume that the deceleration parameter, $q_0$, is 1/2, thus, $R_{i,z}$ is

$$R_{i,z} = \frac{2c}{(1+z_{i,B})H_0}\left((1+z_{i,B}) - (1+z_{i,B})^{1/2}\right). \quad (4)$$

For a given luminosity, $L_0$, we numerically determine $1 + z_{i,B}$ for each of the Band et al. (1993) baseline bursts using equation (2). The result is 50 spectral shapes corrected for redshift commensurate with an assumed $L_0$.

Norris et al. (1994) used ∼ 130 BATSE bursts assigned to the "bright," "dim," and "dimmest" classes. Since the publication of Norris et al. 1994, another 35 bursts have been accumulated in these classes. For the $j\underline{\text{th}}$ Norris et al. (1994) burst in each class, we estimate its redshift, $z_{j,C}$, from its peak intensity, $P_{j,C}$ for each of the Band et al. (1993) spectra (characterized by $\phi_i$, $z_{i,B}$, $L_0$):

$$P_{j,C} = \frac{L_0}{\int_{30}^{2000} E\phi_i\left(\frac{E}{1+z_{i,B}}\right)\frac{dE}{1+z_{i,B}}} \frac{\int_{50}^{300} \phi_j\left(\frac{1+z_{j,C}}{1+z_{i,B}}E\right) dE}{4\pi R_{j,z}^2}, \qquad (5)$$

where $P_{j,C}$ is from the BATSE 2B catalog (Meegan et al. 1994) and the subscript, $C$, refers to the brightness class (i.e., dimmest, dim, bright). We include $(1+z_{j,C})/(1+z_{i,B})$ in this equation to blueshift the observed spectral shape back to $z = 0$ (i.e., divide by $1 + z_{i,B}$) and then redshift the spectrum to a distance of $z_{j,C}$ as seen by us.

Note that we have parameterized the observed peak intensity in terms of the comoving distance. The luminosity distance ($D_L = (1 + z)R_z$) is appropriate for systems measuring the total bolometric flux whereas GRBs observations involve measuring the peak intensity in a finite bandpass. The usual bolometric result can be recovered from equation (5). In other applications, it is assumed that one knows the intrinsic spectrum, that is, $z_{i,B} = 0$. Converting the number spectrum in the numerator of equation (5) (i.e., $\phi(E)$), into an energy spectrum by multiplying by $E$ and integrating over all energies gives the total observed flux:

$$\frac{L_0 \int_0^\infty E\phi[(1+z)E]\,dE}{\int_0^\infty E\phi(E)\,dE\ 4\pi R_z^2} = \frac{L_0}{4\pi(1+z)^2 R_z^2} = \frac{L_0}{4\pi D_L^2} \ .$$

From the 2B catalog we obtained intensity values ($P_{j,C}$ defined over 256 ms) for 38 of the 45 bursts in the bright class, 37 of the 63 dim bursts, and 40 of the 59 dimmest bursts. The derived redshifts from equation (5) for each of the 38 bright bursts for each of the 50 Band et al. (1993) spectra results in 1900 individual redshifts, $z_{j,\text{bright}}$, which are averaged. We repeat this process for different luminosities and for the BATSE dim and dimmest bursts. PVO events make up a fourth class of events even brighter than those seen in BATSE. For PVO we use the Band et al. (1993) spectra and the typical intensity for a bright PVO event, 200 photons cm$^{-2}$ s$^{-1}$. Figure 1a shows the relationship between standard candle luminosity and redshift for the four brightness classes. Also shown on the right hand axis is the comoving distance from equation (4).

The resulting average of 1900 redshifts, $z_C$, accounts for the intrinsic spread of spectral shapes, the spread of intensities within a class, and the fact that the bursts that define the spectral shape are not at $z = 0$. For example, at a luminosity $L_0 = 6.3 \times 10^{51}$ erg s$^{-1}$, the average of the 1900 redshifts derived from the bright class is $z_{\text{bright}} \simeq 1.1$, although there are individual examples as low as 0.31 and as high as 2.7. The root mean square of the distribution of $z_{j,\text{bright}}$ is 0.41, which we assume is an estimate of the intrinsic spread of $z_{\text{bright}}$. The variation in $z_{\text{bright}}$ comes mostly from the variation in $P_{j,\text{bright}}$, rather than due to the variation in the shape of the Band et al. (1993) spectra. Thus, we take as a conservative estimate that the number of independent events is 38 (as opposed to 50 or 1900) and the mean average $z_{\text{bright}}$ for $L_0 = 6.3 \times 10^{51}$ erg s$^{-1}$ is therefore found to be $1.1 \pm 0.07$ (i.e., $\pm 0.41/\sqrt{38}$). Similarly, for the same luminosity, $z_{\rm dim} = 4.4 \pm 0.18$ and $z_{\rm dimmest} = 5.2 \pm 0.20$. Note these error bars only reflect the uncertainty in the mapping of an assumed luminosity to a redshift; they do not represent an uncertainty of a determined value of $z_{\rm bright}, z_{\rm dim}$, or $z_{\rm dimmest}$. These uncertainties are typical for the curves in Figure 1a.

## 3. ENERGY CORRECTION

Time dilation due to the expansion of the universe is counterbalanced by the narrowing of the time history profile at higher energy: the full width half maximum (FWHM) of a pulse as seen in a high-energy bandpass is less than that of a low energy bandpass even without any redshift (Fishman et al. 1992, Link et al. 1993). Thus, the relationship between time dilation and intrinsic luminosity must include a correction factor to account for the narrowing of time structure with redshift. Let the dilation factor $y$ account for the fact that both the bright bursts are not at $z = 0$ and the dimmest bursts are redshifted at some $z_{\rm dimmest}$:

$$y = \frac{1 + z_{\rm dimmest}}{1 + z_{\rm bright}}. \qquad (6)$$

Notice that $y$ is similar to the spectral correction factor in equation (5) which effectively blueshifted the observed spectrum $\phi_i(E)$ to $z = 0$ and redshifted the spectrum to a distance $1 + z_{j,C}$. Figure 1b shows three different $y$ factors each based on a pair of redshifts from Figure 1a as a function of luminosity $L_0$.

The time dilation ($S_{B-D}$) of the dimmest bursts relative to the bright is directly related to $y$. Let $W(E_{L,k}, E_{U,k})$ be the average FWHM of the time structure in the $k\underline{\rm th}$ BATSE bandpass (from $E_{L,k}$ to $E_{U,k}$ keV) of bright bursts which occurs at an average redshift of $z_{\text{bright}}$. If that burst occurred further away, say at redshift $z = z_{\text{dimmest}}$, the recorded width in the $k\underline{\text{th}}$ bandpass can be derived from the bright bursts using the energy range $yE_{L,k}$ to $yE_{U,k}$. Let $W(yE_{L,k}, yE_{U,k})$ be the average FWHM of a pulse as would be seen in the energy range $yE_{L,k}$ to $yE_{U,k}$ keV. The observed time dilation of a dimmest burst relative to a bright burst is:

$$S_{B-D,k} = y \frac{W(yE_{L,k}, yE_{U,k})}{W(E_{L,k}, E_{U,k})}. \tag{7}$$

We refer to $W(yE_{L,k}, yE_{U,k})/W(E_{L,k}, E_{U,k})$ as the $W$-correction and it accounts for the tendency for higher energy peaks to be narrower in individual bursts (Link, Epstein, & Priedhorsky 1993). To develop the width function dependence on $y$ (i.e., $W(yE_L, yE_U)$), we fit the observations of average pulse widths used by Norris (1994) and Davis et al. (1994) to a model of how the pulse width varies with energy. We assume that the time structure can be separated from the spectral shape, that is, $\phi(t, E) = A(t, E)\phi(E)$. Let $R_k(E)$ be the effective area function for the $k\underline{\text{th}}$ BATSE bandpass, then the time history observed in the $k\underline{\text{th}}$ energy band, $H_k(t)$, can be modeled as:

$$H_k(t) = \int_0^\infty R_k(E) A(t, E) \phi(E) dE, \tag{8}$$

where the energy bands of $R_k(E)$ are 25 keV to 57 keV, 57 keV to 115 keV, 115 keV to 320 keV, and 320 keV to $\sim$ 1000 keV, corresponding to $k = 1, 2, 3, 4$, respectively. Combined channels, such as 25 to 115 keV, use the sum of the respective effective area functions (e.g., $R_{1+2}(E) = R_1(E) + R_2(E)$). To construct a profile, we assume a scaling factor, $A(t, E)$, that is based on

an exponential shape:

$$A(t, E) = \exp\left(\frac{-t}{\tau(E)}\right). \tag{9}$$

The spectral shape is not just $\phi(E)$ due to the energy dependence in $A(t, E)$. Variations in the spectral shape are not important because they have a minimal effect on weighting $A(t, E)$ in equation (8). In fact, the variations due to $A(t, E)$ are necessary because without spectral evolution through the pulse one cannot obtain a narrowing at higher energies. We have found that the $A(t, E)$ dependence mimics both the changing intensity within a peak and the hard to soft evolution. The decay time of the pulse at a given energy, $\tau(E)$, is assumed to have the form

$$\tau(E) = S_1 \left(\frac{E}{100}\right)^{S_2}, \tag{10}$$

where $S_1$ and $S_2$ are unknown parameters and $E$ is in units of keV. The motivation for using a power law is twofold. First, the auto-correlation function tends to more accurately measure width than the average pulse shape and it shows a dependence similar to a power law (Fenimore et al. 1995a). Second, the power law fits the data very well (to within tens of milliseconds) and agrees well at energies where it can be estimated (see Figure 2). In contrast, a linear function for the time decay (i.e., $\tau(E) = S_1 E + S_2$) was unworkable. In addition, a power law function with a breakpoint does not give a significant improvement in the fit.

We construct a composite profile by averaging the time histories over a range of spectra observed by BATSE:

$$H_k(t) = \left(\frac{1}{n}\right) \sum_{i=1}^{n} \int_0^\infty R_k(E) A(t, E) \phi_i(E) dE, \tag{11}$$

where each $\phi_i(E)$ corresponds to different Band et al. (1993) spectra ($n = 54$). The FWHM of $H_k(t)$ is $W(E_{L,k}, E_{U,k})$.

Since the profiles are asymmetric about the peak, we decompose each pulse profile into a rise and fall side. The widths (in seconds) are found from the profiles as (rise/fall): 0.22/0.44, 0.17/0.32, 0.13/0.27, and 0.08/0.18 for BATSE channels 1, 2, 3, and 4, respectively. Using a method of least squares, we fit the four profile widths generated from equation (11) for both sides to these widths and find the four best fit constants for $\tau(E)$. With the constants $S_{1,\text{rise}} = 0.22$, $S_{1,\text{fall}} = 0.45$, $S_{2,\text{rise}} = -0.40$, and $S_{2,\text{fall}} = -0.39$, the residual (in seconds) between observed and calculated widths (rise/fall) are: $1 \times 10^{-5}/9 \times 10^{-3}$, $-7 \times 10^{-4}/-2 \times 10^{-2}$, $1 \times 10^{-3}/5 \times 10^{-3}$, and $-9 \times 10^{-4}/6 \times 10^{-3}$ for channels 1, 2, 3, and 4, respectively. Although no goodness of fit can be presented here because we do not know the uncertainties in the pulse widths, the calculated widths always match the observed to better than a few tens of milliseconds. These residues of the fit are certainly within the uncertainties of the widths themselves.

When the spectrum is redshifted, the photon energy in the equation for $H_k(t)$ is shifted by $y$ in those terms that are generated at the burst:

$$H_{y,k}(t) = \left(\frac{1}{n}\right) \sum_{i=1}^{n} \int_0^\infty R_k(E) A(t, yE) \phi_i(yE) dE. \qquad (12)$$

Notice that the response matrix is unaffected so the only effect on the time history is due the to energy redshift in $\phi(yE)$ and $A(t, yE)$. Here, the FWHM of $H_{y,k}(t)$ is $W(yE_{L,k}, yE_{U,k})$. Figure 2 shows $W(yE_L, yE_U)/W(E_L, E_U)$ based on the best fit width functions for BATSE channels 1, 1+2, and 3. These width corrections are based on the average pulse shape so, technically, should only be used to correct time dilation found from the average pulse shape test. We believe they can also be used for any test that involves linear combinations of peaks (such as the peak-align test) and senses time structure the order of a few seconds in a burst. Also shown in Figure 2 is $W(yE_L, yE_U)/W(E_L, E_U)$ for the auto-correlation test (Fenimore et al. 1995b) which depends quadratically on the counts.

One verification that the $W$-correction is roughly correct can be seen by comparing the ratios of widths constructed from our fitted functions at a particular $y$ to the ratio of observed pulse widths for energy bands where $(yE_{L,k}, yE_{U,k})$ is roughly $(E_{L,j}, E_{U,j})$. There are two such cases for the average pulse energy correction and one for the auto-correlation function. For example, from the average fitted pulses, $W(E_{L,1}, E_{U,1}) = 0.65$ sec and $W(E_{L,2}, E_{U,2}) = 0.49$ sec. Our fitting gives $W(2E_{L,1}, 2E_{U,1}) = 0.767 W(E_{L,1}, E_{U,1})$, which closely matches expectation (0.49/0.65) since a $y$ factor of 2 approximately maps the energy range of the first BATSE channel (25-57 keV) to the energy range of the second channel (57-115 keV). In Figure 2, the solid square represents the observed ratio of $W(E_{L,1}, E_{U,1})$ to the observed width of $W(E_{L,2}, E_{U,2})$ which does agree with the calculated $W(yE_{L,1}, yE_{U,1})/W(E_{L,1}, E_{U,1})$ when $y = 2$. The solid triangle is the observed ratio of $W(E_{L,4}, E_{U,4})$ to $W(E_{L,3}, E_{U,3})$, which agrees well with $W(yE_{L,3}, yE_{U,3})/W(E_{L,3}, E_{U,3})$ when $y = 2.8$. This gives us confidence that our width functions are valid at least to $y \sim 3$. Unfortunately, there is not such a point where one can check the width function for the 25 to 100 keV bandpass that was used in most of the time dilation tests of Norris et al. 1994 and Davis et al. 1994. Based on the points where we can check them, our width functions seem conservative: the functions actually give a smaller correction than if some interpolation of the points were used. It is more difficult to establish a width function for the auto-correlation function since the auto-correlation function depends quadratically on the counts. The open symbol in Figure 3 is a point at which we know the width of the auto-correlation function. It, too, is valid to $y \sim 3$.

## 4. RELATING TIME DILATION TO DISTANCE

Figure 1a uses the observed intensities to give GRB distances if $L_0$ is known. Since we have averaged over spectral shape and burst intensity, Figure 1a only depends on our choice of $q_0$ and $H_0$. Figure 1b is obtained directly from Figure 1a. Figure 1c uses the $W$-correction from Figure 2 with Figure 1b to obtain a relationship between the observed time dilation and $L_0$. Thus, the observed time dilation gives $L_0$ which can be used in Figure 1a to give the distance to cosmological GRBs.

Figure 1c is for the BATSE bandpass 25-115 keV. Two different functions have been plotted corresponding to the bright-dimmest and bright-dim $y$ factors from Figure 1b. Figure 1c assumes the energy-width relationship that was used for most of the tests (i.e., 25 to 115 keV) and, technically, it can only be used for the average pulse width test. However, given the similarity between the various functions in Figure 2, we believe Figure 1c can be used to estimate the standard candle luminosity for other tests that target the same time scale within the GRBs, that is, a few seconds. Figures 3a and 3b show similar time dilation for the BATSE energy bandpass 25-57 keV and 115-320 keV. Figure 3c uses the energy correction from the average auto-correlation function (cf. Fig. 2) so provides an accurate mapping of time dilation to luminosity for that test.

Table 1 summarizes the various tests (first column) that have been used to detect time dilation. The second column gives the measured time dilation. The next two columns ignore the $W$-correction: the observed time dilation from Table 1 is set equal to $y$ in Figure 1b and mapped to an acceptable standard candle luminosity using the "Dimmest-Brt" curve. This range is reported in column 3 of the table and used with the "Dimmest" curve of Figure 1a to determine the corresponding range of $z$ (column 4). The $\pm 1\sigma$ ranges for $L_{50}$ ($= L_0 \times 10^{-50}$) are shown in Figure 1b for those tests where a confidence region has been quoted. The solid horizontal lines at the bottom of Figure 1b give the acceptable range of the standard candle luminosity for the fluence/peak test, the average pulse width test, the average counts in a peak test, the total counts test, the duration test, and the average auto-correlation test (from bottom to top, respectively). Even without the $W$-correction, the observed time dilation indicates a larger standard candle luminosity than consistent with the log $N$-log $P$ distribution (the dotted horizontal line). The best tests (duration and auto-correlation) have an average standard luminosity of $\sim 20 \times 10^{50}$ erg s$^{-1}$ if the $W$-correction is ignored. In our comparison to the log $N$-log $P$ distribution (section 5), we will use $2 \times 10^{51}$ erg s$^{-1}$ as an example of the luminosity implied by time dilation tests without an $W$-correction. Note that the $\pm 1\sigma$ range on $z_{\text{dimmest}}$ (column 4) is usually quite large compared to $z_{\text{dimmest}}$ because of the steepness of Figure 1c. This supports our position that one should quote time dilation results in terms of the standard candle luminosity rather than the $z$ at some ill-defined threshold.

Energy corrections *are* clearly required. Columns 5 to 8 in Table 1 incorporate the $W$-correction. Some tests (wavelet power, total counts, duration) sense time dilation on long time scales within the bursts, and it is less likely that Figure 2 is an appropriate $W$-correction. These tests cannot be mapped to a standard candle luminosity until their energy-width relationship have been determined. To include the $W$-correction, the observed time dilation from Table 1 is set equal to the ordinate of Figure 1c (3c for the autocorrelation test) and mapped to an acceptable range of standard candle luminosity using the "Dimmest-Brt" curve. This range is reported in column 6 of Table 1 and used with the "Dimmest-Brt" curve of Figure 1b to obtain the corresponding $y$ (reported in column 5). The luminosity range is used with the "Dimmest" curve of Figure 1a to obtain the corresponding $z$ and comoving distance (i.e., eq. [4]) which are reported in columns 7 and 8 of Table 1. The $\pm 1\sigma$ range for the tests are shown in Figure 1c as horizontal solid lines. From bottom to top, the lines represent the fluence/peak test, the pulse width test, the counts in a peak test, and the auto-correlation test. These tests imply that the standard candle luminosity consistent with the time dilation when one includes the $W$-correction is the order of $10^{52}$ erg s$^{-1}$. We will use $10^{52}$ erg s$^{-1}$ in our comparisons to the log $N$-log $P$ distribution (see section 5).

If the time dilation is due to the expansion of the universe, then it must occur equally in all time scales. Indeed, all the tests for which an $W$- correction can be estimated are consistent. However, the resulting standard candle luminosity is much larger than found from the log $N$-log $P$ distribution.

## 5. COMPARISON TO LOG N-LOG P

One goal of this paper is to investigate the consistency between the distance commensurate with the time dilation and the distance commensurate with the log $N$-log $P$ distribution. There have been a fair number of log $N$-log $P$ studies, and they have not arrived at consistent results. Most papers have concluded that $z_{max}$ is about unity. However, even if one just considers results from the same assumed spectral shape, the standard candle luminosity concluded by these papers vary by two orders of magnitude from $< 4 \times 10^{49}$ erg s$^{-1}$ (Piran 1992) to $4 \times 10^{51}$ erg s$^{-1}$ (Dermer 1992). An additional problem in comparing results arises from using $z_{max}$ since $z_{max}$ only has a meaningful value if the threshold it corresponds to is quoted and the reader is told how that threshold maps into luminosity. We recommend that studies characterize their results in terms of the standard candle luminosity in a specified energy range rather than $z_{max}$. We have used the energy range 30 to 2000 keV. Most papers have assumed power law number spectra ($E^{-\alpha}$) and the resulting $z_{max}$ and $L_0$ varies strongly with $\alpha$. For example, just varying $\alpha$ from 1.5 to 2.5, Wickramasinghe et al. 1993 found $z_{max}$ could vary from 0.5 to 4.0. Associated with these variations with $z_{max}$ were large variations in $L_0$. Given the strong dependence of $L_0$ on the assumed spectra shape, we suggest that the spectral shape must be accurately modelled. Only Dermer 1992 and Fenimore et al. 1993 have used spectral shapes with curvature similar to that observed. However, these studies concluded luminosities that were different: Dermer 1992 found $L_0 = 4 \pm 2 \times 10^{51}$ erg s$^{-1}$ with $z_{max} = 1.2$ and Fenimore et al. 1993 found $L_0 = 5.2^{+1.4}_{-1.3} \times 10^{50}$ erg s$^{-1}$ in the bandpass 30 to 2000 keV corresponding to $z = 0.79 \pm 0.05$ for bursts with peak intensities of 0.7 photons cm$^{-2}$ s$^{-1}$. Dermer 1992 compared models that ignored threshold effects to the $V/V_{max}$ of the first 126 bursts observed by BATSE. Threshold effects can only be ignored if the log $N$-log $P$ distribution is a power law (Band 1993), perhaps contributing to the discrepancy. The Dermer 1992 confidence region is large, and it appears that it could include the Fenimore et al. 1993 value at a $2\sigma$ level. We note that the Dermer confidence region was based on only 126 events, whereas the Fenimore et al. region used bursts from a much larger dynamic range, equivalent to 4200 BATSE bursts.

We will use the Fenimore et al. analysis. This is a crucial decision since we will later conclude that the amount of the time dilation is too large to be consistent with the distance determined from the log $N$-log $P$ distribution. It must be recognized that we are not using the log $N$-log $P$ study that gave the largest distance (Dermer 1992) because of the reasons discussed above. We are confident that studies that used only power law spectra (Loredo and Wasserman 1994, Mao and Paczyński 1992, Petrosian, Azzam, and Meegan 1994, Piran 1992, and Wickramasinghe et al. 1993 as well as our earlier work, Fenimore et al. 1992) are undependable. The Fenimore et al. 1993 log $N$-log $P$ study combined events from BATSE and PVO. Due to its small size, PVO only saw the brightest events, but PVO had a longer lifetime ($\sim 14$ yr), larger on-time (90%), and a larger field of view ($4\pi$ since PVO had very little Venus blockage). It will take BATSE $\sim 30$ more years to observe as many bright bursts as PVO. We apply a uniform selection criterion for the included events and accommodate the differences in instrumental response in calculating intensities. The threshold effects are accounted for. In Fenimore et al. 1993, we assumed that the intrinsic spectral shape is $\phi(E) = E^{-1} \exp(-E/350)$ such that

$$L_0 = \int_{30}^{2000} E \phi_0 \phi(E) dE \qquad (13)$$

and related $z$ to intensity by

$$P_{j,C} = \frac{L_0}{\int_{30}^{2000} E \phi(E) dE} \frac{\int_{100}^{500} \phi[(1+z_{j,C})E] dE}{4\pi R_{j,z}^2} \quad . \qquad (14)$$

(Note in Fenimore et al. 1993 we used an energy range that was common to both BATSE and PVO, 100 to 500 keV.) We fit the observed number of bursts in a range of peak intensities to

$$\Delta N(P_1 \text{ to } P_2) = \int_{R_1}^{R_2} \frac{\rho}{1+z} dV \quad . \qquad (15)$$

Here, $\rho$ is the rate-density of events per comoving volume (bursts yr$^{-1}$ Gpc$^{-3}$). In contrast to steady sources (such as galaxies), equation (15) requires the extra $1 + z$ in the denominator because the time between bursts is dilated with the expansion of the universe. For $q_0 = 1/2$, $dV$ has the simple form

$$dV = 4\pi r^2 dr. \qquad (16)$$

Such an analysis effectively assumes that we know the intrinsic spectral shape (i.e., at $z = 0$) when, in fact, we have only observed it for the bright bursts which would be at a substantial $z$ if $L_0$ is large. Thus, in this paper we make two improvements to our previous log $N$-log $P$ study. (These improvements have not been utilized in any other study that we are aware of.) First, when we do the fitting using equation (15) we average over a set of spectral shapes that span the parameter space for GRBs, that is, we use the 50 Band et al. 1993 spectra in equations (1) and (5). Second, we introduce a new parameter, $z_B$, the redshift of the events that define the spectra shape: compare equation (1) with (13) and (5) with (14). We use an additional relationship (eq. 2) with additional information ($P_{i,B}$) to determine $z_B$. (When using eq. [5] for the log $N$-log $P$ studies, we use an integration range of 100 to 500 keV in the numerator; and when we use eq. [5] for Fig. 1 or 3, we use a range of 50 to 300 keV.)

The histogram in Figure 4 shows the log $N$-log $P$ distribution of the combined, uniformly selected events from PVO and BATSE. Table 2 provides the intervals ($P_1$ and $P_2$) and the number of observed events ($\Delta N_{obs}$) used for the fitting of the differential log $N$-log $P$ to equation (15). The fitting used a $\chi^2$ statistic defined as

$$\chi^2 = \sum \frac{\left(A_s \Delta N_{obs} - \Delta N(P_1 \text{ to } P_2)\right)^2}{A_s \Delta N(P_1 \text{ to } P_2)}, \qquad (17)$$

where $A_s$ normalizes the observed number of events in a photon s$^{-1}$ cm$^{-2}$ range to events yr$^{-1}$ $(4\pi)^{-1}$. For PVO, $A_s = 0.09072$ and for BATSE, $A_s = 4.237$ (see Fenimore et al. 1993). We did not use data below where either instrument has threshold effects. Note that this uses the model rather than the observations to estimate the variance on the observations. Table 2 can be used to fit any model to the combined BATSE-PVO log $N$-log $P$. The results of the re-analysis is $L_0 = 4.6^{+0.90}_{-0.65} \times 10^{50}$ erg s$^{-1}$ in the bandpass 30 to 2000 keV with $\rho = 24$ Gpc$^{-3}$ yr$^{-1}$, rather than $5.2 \times 10^{50}$ erg s$^{-1}$ as reported in Fenimore et al. 1993. The solid line is the expected distribution given a best-fit standard candle luminosity of $4.6 \times 10^{50}$ erg s$^{-1}$ the corresponding $\chi^2$ is 9.1 with 9 degrees of freedom. The dashed line in Figure 4 represents the expected log $N$-log $P$ distribution of events given a standard candle luminosity of $10^{52}$ erg s$^{-1}$, the value consistent with the time dilation tests; it poorly matches the observed distribution from BATSE and PVO ($\chi^2$ is 510). The dotted line is for $L_0 = 2 \times 10^{51}$ erg s$^{-1}$, the standard candle luminosity obtained for time dilation without any $W$-correction. It, too, poorly matches the observed BATSE-PVO log $N$-log $P$ distribution ($\chi^2$ is 60).

Perhaps evolution can explain the discrepancy (Norris et al. 1995). The expected log $N$-log $P$ distribution can be modified by either number evolution (i.e., $\rho = \rho(z)$) or luminosity evolution (i.e., $L = L(z)$). Spectral evolution has little effect on the expected distribution. We use the functional shapes $L = L_0(1+z)^{p_L}$ and $\rho = \rho_0(1+z)^{p_\rho}$ which were assumed by Fenimore et al. 1993. However, since $N$ scales as $\rho L_0^{3/2}$ for most of the parameter range, we effectively fit with $(1+z)^{p_\rho + \frac{3}{2}p_L}$. In Figure 5a, we show the evolution necessary ($p_\rho + \frac{3}{2}p_L$) to achieve agreement with the BATSE-PVO log $N$-log $P$ as a function of standard candle luminosity. The fits are unacceptable for $p_\rho + \frac{3}{2}p_L$ greater than $\sim 2.4$. Of course, other functional forms for the evolution might be able to provide acceptable fits. The required evolution is quite strong and a successful cosmological GRB model must ac-commodate it. For example, assume $p_\rho + \frac{3}{2}p_L = 2.4$ even though it provides insufficient evolution to obtain a best fit (cf. Fig. 5a). The time dilation requires $z$ to be at least 6 (cf. Table 1). Thus, if one has density evolution, the density of progenitors must be a factor of $\sim 100$ higher at the distance of the dimmest bursts. It requires a factor of $\sim 10$ higher density at the distance of the bright BATSE bursts. In the context of a colliding neutron star model for GRBs, this would require that neutron star production was much higher in the distant past (which is not inconceivable). Alternatively, if luminosity evolution occurs, $p_L$ is 1.6 and the objects must be $\sim 25$ times brighter at the distance of the dimmest events. Since neutron stars have a relatively small range of masses, it seems unlikely that colliding neutron stars would release more energy just because the universe is younger. However, the process that converts the released energy to gamma rays is uncertain and could depend on epoch. Epstein et al. (1993) suggested that GRBs are upscattered ambient photons from colliding stars near an AGN. AGNs are brighter in the distant past so, perhaps, the reason GRBs have luminosity evolution is that there are more ambient photons in the vicinity of the colliding neutron stars.

Perhaps more troublesome is that the evolution required for consistency between the time dilation and the log $N$-log $P$ distribution coincidentally produces a $-3/2$ power law in the PVO log $N$-log $P$ distribution. Figure 5b gives the $\chi^2$ for fitting just the 6 PVO bins above 20 photons cm$^2$ s$^{-1}$ and the BATSE bin above 7.08 photon cm$^2$ s$^{-1}$ (see Table 2) with equation (15). This is the region of the log $N$-log $P$ distribution that follows a -3/2 power law. Low luminosities give an acceptable $\chi^2$ of 5.0 (with 5 degrees of freedom).

The $\chi^2$ grows quickly for luminosities greater than $10^{51}$ erg s$^{-1}$. For the luminosity consistent with time dilation (with the $W$-correction), the $\chi^2$ has grown to 30. Evolution is required to obtain an acceptable fit in the -3/2 portion of the distribution. Thus, the shape of the dashed curve in Figure 4 is inconsistent with the $-3/2$ portion of the log $N$-log $P$ distribution. Strong evolution can modify the shape to agree with the -3/2 portion but then it must be considered a coincidence that the log $N$-log $P$ distribution has a portion that appears to be from a homogeneous distribution. The chances that such a coincidence would occur is roughly the probability of obtaining a $\chi^2$ of 30 with 5 degrees of freedom.

In previous studies it was usually assumed that we knew the intrinsic shape of the spectrum (i.e., eqs. [13] and [15] were used rather than eqs. [1] and [5]). The inclusion of $z_B$ improves the comparison between the time dilation and the log $N$-log $P$ distribution. If $z_B$ is assumed to be zero, then Figure 1 would actually give even a larger $L_0$ and $z_{\rm dimmest}$. The inclusion of the $z_B$ terms moderates variations in $L_0$. Even though $z_{\rm dimmest}$ is quite large ($> 6$), the corresponding $L_0$ does not vary much from a low $z$ value. The reason is that $(1 + z_{\rm dimmest})/(1 + z_{\rm bright})$ is limited by the dynamic range of the observed GRBs peak intensities, which is small (about a factor of 30 for BATSE).

## 6. OTHER COSMOLOGIES

The luminosity implied by the time dilation has many fewer faint events than observed (cf. Fig. 4). Strong evolution will allow some reconciliation of the log $N$-log $P$ distribution and the time dilation. Another possible way to reconcile the difference is to assume different cosmologies. Cosmologies with low $q_0$ will produce relatively more faint events for the same standard candle luminosity. However, we will show that small $q_0$ requires a much larger luminosity such that the time dilation is not consistent with the log $N$-log $P$ distribution. For $q_0$ near zero, equation (3) becomes

$$R_{i,z} = \frac{cz}{H_0(1+z)}(1 + \frac{z}{2}). \tag{18}$$

Using equation (18) in equations (2) and (5), we produce Figure 6 which is similar to Figure 1. The redshift, $z$, varies more slowly with luminosity than in the case of $q_0 = 1/2$. At a given luminosity (say, $10^{52}$ erg s$^{-1}$), The $z$ of the dimmest events is not much different than the $z$ of the bright events. Thus, $y$ is relatively smaller for $q_0 = 0$ compared to $q_0 = 1/2$. If one ignores the $W$-correction and identifies the observed time dilation of 2.0 with $y$ in Figure 6b, then the required luminosity is $\sim 2.8 \times 10^{52}$ erg s$^{-1}$. When $q_0 \neq 1/2$, $dV$ in equation (15) takes the form

$$dV = 4\pi\left(\frac{c}{H_0}\right)^3 \frac{\left[q_0 z + (1-q_0)\left(1 - (1+2q_0 z)^{1/2}\right)\right]^2}{q_0^4(1+z)^3(1+2q_0 z)^{1/2}} dz , \tag{19}$$

(see eq. [2.56] in Kolb & Turner 1992 ). Using equation (19) in equation (15) gives a $\chi^2$ of 314 for $L_0 = 2.8 \times 10^{52}$ erg s$^{-1}$. Since $y$ is always small (due to the relatively small differences in $z$ in Fig. 6a), there are no reasonable luminosities that produce a time dilation as large as observed (Fig. 6c).

The other extreme case is to assume that $q_0 = 1.0$. Although unlikely, it does allow better agreement between the log $N$-log $P$ distribution and the time dilation. In this case, equation (3) becomes

$$R_{i,z} = \frac{cz}{H_0(1+z)}. \tag{20}$$

For $q_0 = 1$, we find that the log $N$-log $P$ distribution is best fit by a luminosity of $3.7 \times 10^{50}$ erg s$^{-1}$, with $\chi^2 = 9.1$ for 9 degrees of freedom. Using equation (20) in equations (2) and (5), one can repeat the analysis of Figure 1 and determine the standard candle luminosity that is consistent with the observed time dilation. One finds for $q_0 = 1$ that a luminosity of $7 \times 10^{50}$ erg s$^{-1}$ has a time dilation of 2.0 with no $W$-correction and a value of $2.75 \times 10^{51}$ erg s$^{-1}$ including the $W$-correction. The assumption of $q_0 = 1$ allows the dimmest BATSE events to be at a lower $z$, about 5. Figure 7 is similar to Figure 4 except it assumes $q_0 = 1$. A luminosity of $7 \times 10^{50}$ erg s$^{-1}$ fits the log $N$-log $P$ distribution with a $\chi^2$ of 18 and a luminosity of $2.75 \times 10^{51}$ erg s$^{-1}$ fits with a $\chi^2$ of 180. Thus, the luminosities that fit the time dilation are closer to that which fits the log $N$-log $P$ distribution but are still inconsistent.

In Figure 5, the dotted curves are for $q_0 = 1$. At $2.75 \times 10^{51}$ erg s$^{-1}$, substantial evolution is required ($p_\rho + \frac{3}{2} p_L = 3.2$, Fig. 5a). If one just fits to the -3/2 portion of the log $N$-log $P$ distribution, the luminosity consistent with the time dilation fits with a $\chi^2$ of $\sim 12$ (see Fig. 5b). Since the best fit $\chi^2$ for the -3/2 portion is 5 and there are two fit parameters, the time dilation luminosity is approximately at the $2\sigma$ contour of acceptable log $N$-log $P$ luminosities. Thus, one can reconcile the time dilation and log $N$-log $P$ distribution if $q_0 = 1$ and there is substantial evolution ($p_\rho + \frac{3}{2} p_L = 3.2$). In that case, it is a $2\sigma$ coincidence that we see a -3/2 power law at high intensities.

## 7. DISCUSSION

All tests show time dilation of $\sim 2$ between the BATSE bright bursts and the dimmer bursts (Norris 1994, Norris et al. 1994, Davis et al. 1994, Fenimore et al. 1995b, although see Mitrofanov et al. 1994). The key question is what distance does a time dilation of 2 imply and is it consistent with the log $N$-log $P$ distribiution. Norris et al. (1995) estimates $z_{dimmest}$ based on a consistency argument. Norris et al. (1995) uses $\mathcal{S}_{B-D} = 2 \sim y$ as a first order approximation to estimate the $W$-correction and obtains a $W$-correction for the average pulse width test equal to 0.85 which corrects $y$ to 2.35. The Norris et al. (1995) conversion from $y$ to $z$ uses equation (6) and assumes $z_{Brt} = 0.3$ (the log $N$-log $P$ value for the bright BATSE events) and concluded that $z_{dimmest} \sim 2$. This is close to being self-consistent, for $z_{dimmest} = 2$, $z_{brt}$ is actually $\sim 0.6$. But, as $z_{brt}$ get larger, the $W$-correction and $z_{dimmest}$ get larger causing $z_{brt}$ to be larger. Thus, Norris et al. (1995) obtained a low value for $z_{dimmest}$ primarily because $z_{brt}$ was assumed to be 0.3. The purpose of this paper is to formulate the relationship between time dilation and distance including all known effects. Rather than an iterative process, we calculates the function in Figure 1c which maps $\mathcal{S}_{B-D}$ directly to $L_0$ and, therefore, $z_{dimmest}$ and $y$. A $\mathcal{S}_{B-D}$ of 2 gives a $y$ of 3 and the $W$-correction is $\sim 0.68$. We use a different conversion from $y$ to $z$ then did Norris et al. (1995). For example, our conversion recognizes that the intensity and spectral shape used to define $z$ vs. $L_0$ were not from $z = 0$ (note the $z_b$ terms in the eqs. [1] to [5]). Thus, even if one accepts the Norris et al. (1995) $y$ value of 2.35, the corresponding $z_{dimmest}$ found by Figure 1 is 4, not 2. When one uses Figures 1c and 1a, the $z_{dimmest}$ corresponding to, for example, $\mathcal{S}_{B-D} = 2^{+0.50}_{-0.25}$ is $5.5^{+8.2}_{-2.4}$ (see Table 1). Thus, for the same observation, different analysis methods have resulting $z_{dimmest}$'s that vary from 2 to 5.5. At a low value of $z = 2$, the log $N$-log $P$ distribution result ($z = 0.8$, Fenimore et al. 1993) might be within the statistics or modest evolution could produce consistency (Norris et al. 1995). However, this distance scale for the dimmest events was arrived at by assuming that the bright events were at the $z$ consistent with the log $N$-log $P$ distribution. Although reasonable if $z_{brt}$ is small, a more detailed analysis is required if $z_{dimmest} \gtrsim 0.5$. At $z = 5.5$, even extreme evolution has difficulty obtaining consistency (see section 5).

Even if the $W$-correction term is ignored, time dilation of $\sim 2$ implies $z_{\mathrm{dimmest}}$ is $\sim 2.5$ or a standard candle luminosity of $2 \times 10^{51}$ erg s$^{-1}$. This result depends only on the conversion between peak intensity and standard candle luminosity (i.e., Fig. 1a). To be consistent with the log $N$-log $P$ distribution, such a high luminosity requires some evolutionary effects. When the required $W$-correction is included, an observed time dilation of $\sim 2$ actually maps into a $z_{\mathrm{dimmest}}$ of $\gtrsim 6$ that is, a large standard candle luminosity ($10^{52}$ erg s$^{-1}$) requiring substantial evolution to be consistent with the log $N$-log $P$ distribution. This result depends on our estimate of the $W$-correction term in equation (7), which is shown in Figure 2. A time dilation of 2 implies that $y$ is about 2.9. The $W$-correction term is reasonably well determined to at least $y$ equal to 3 (see solid points in Fig. 2). Thus, we do not think that the $W$-correction term is falsely implying a large $L_0$. When $y \sim 2.9$, the average redshift ($z$) of the BATSE bright bursts is $\sim 1.2$ and the average redshift for the dim events is $\sim 5.6$ (cf. Fig. 1a). Thus, when there is net time dilation of $\sim 2$, the time dilation due to the expansion of the universe is responsible for a factor of 2.9, but that is counteracted by the narrowing of the pulse profiles with energy which contributes a factor of 0.68.

Beaming can reduce the overall energy requirements to be less than $10^{52}$ erg s$^{-1}$. However, beaming does not change our estimate of $z_{\rm dimmest}$; it only scales the reported luminosity by $\Omega/4\pi$ where $\Omega$ is the angular size of the beam. Beaming does not affect our comparison to the log $N$-log $P$ distribution, since $L_0$ scales in the same way in both analyses.

For many models, a $z$ of $> 6$ is uncomfortably large. For example, to be consistent with the neutron star merger model (Mészaros and Rees 1992), one must assume a very early phase with a much higher density of neutron star pairs. Since galaxies are thought to form at about $z \sim 3$, it is not clear why there should be so many neutron star pairs so early. On the other hand, the greater distance would help resolve the "no host" problem. Schaefer (1992) pointed out that small error boxes for bright (nearby) GRBs are devoid of bright extragalactic objects such that M31-like galaxies could not be the host galaxy. Fenimore et al. 1993 showed that it was difficult to reconcile the log $N$-log $P$ distribution with any host galaxy if GRBs followed the luminous mass. Certainly bright AGNs were ruled out. If the greater distance implied by the observed time dilation is correct, then it is not surprising that we do not see the host galaxies. The no-host problem does not exist if GRBs are associated with galactic centers since the number of galaxies (in contrast to the luminous mass) could be very large for small, virtually undetectable, galaxies.

There are three possible explanations that could reconcile the excessive time dilation with the log $N$-log $P$ distribution. First, the time dilation could, indeed, be a result of the expansion of the universe and the bursts come from extreme distances ($z$ beyond 6). The lack of normal galaxies within the error boxes would no longer be a problem, although the bursts would be orphans: no know objects with the same distance scale. Either strong density evolution or luminosity evolution would be required. The luminosity evolution might result from models that utilize the ambient radiation of AGNs (cf. Epstein et al. 1993). Under this explanation, the observed -3/2 power law in the log $N$-log $P$ distribution must be considered a coincidence rather than indicating that some of the bursts are from a nearby, homogeneous population.

Second, the log $N$-log $P$ distribution could indicate a valid distance scale, but, therefore a large fraction of the observed time dilation is intrinsic to the bursts. We tend to favor this explanation since it is probably easy to have intrinsic time stretching of the bursts. The distance indicated by the log $N$-log $P$ distribution implies that the dimmest GRBs should only have a time stretch of $\sim 1.3$ (cf. Fig. 1c) rather than the observed 2. Thus, 65% of the time stretching must be intrinsic in order to be consistent with the log $N$-log $P$ distribution. Under this explanation, it would be difficult to use the time dilation as an argument that GRBs are cosmological since perhaps all of the time dilation could be intrinsic.

Indeed, Band (1994) and Wijers & Paczyński (1994) have argued that such time dilation could easily be intrinsic to GRBs. Clearly, the most suspect assumption that Norris et al. (1994) and we have made is that peak intensity indicates distance. Just based on the variation of peaks within individual bursts, it is reasonable that GRB peak intensities have a very broad distribution even for sources that are at the same distance. If bright bursts intrinsically have shorter time scales, then we could just be seeing the result of a luminosity function. This could arise easily if, for example, fluence ($\sim$ luminosity times duration) was standard rather than peak intensity, see Band (1994) and Wijers & Paczyński (1994). All reports of time dilation have assumed a standard candle luminosity and, if bursts have, in fact, a standard candle fluence, then the reports of time dilation are probably artifacts of the standard candle luminosity assumption. Unfortunately, total fluence is harder to measure and the trigger efficiency for fluence is so poorly known that a standard candle fluence analysis is probably not possible for time dilation.

Third, all of the time dilation tests that we have used (Norris 1994, Norris et al. 1994, Davis et al. 1994, Norris et al. 1995, Fenimore et al. 1995b) are based on the same set of events. Although all efforts to identify systematic effects have been taken and statistical variations accounted for, perhaps there is an unknown systematic effect that gives the appearance of a correlation of intensity and time scale. Not all bursts were used, some had data gaps that made them unusable and these tend to be the longer bursts. The shortest events (less than 1.5 s duration) were omitted, although it is not clear why that would induce a correlation. It is our judgment that systematic effects have been accounted for.

Finally, we make the following recommendations concerning time dila-tion tests. First, one must use $yW(yE_L, yE_U)/W(E_L, E_U)$ to properly relate the observations to a distance scale rather than $1 + z_{\text{dimmest}}$. Although the differences are small they can cause a large effect. Second, for each test, the bright events must be used to provide $W(E_L, E_U)$. Third, $z_{\text{dimmest}}$ is rather nebulous because it depends on the threshold of the instrument. One should quote the equivalent $L_0$ in a specified energy range. Error bars on $z_{\text{dimmest}}$ often are large because $z$ diverges at distances near $2c/H_0$. Fourth, the assumption of simple spectral shapes, such as power laws or thermal bremsstrahlung, is unwarranted. One should use averages over the Band et al. 1993 spectra. And, fifth, one cannot assume that the events that define the spectral shape that are used in either the log $N$-log $P$ studies or the conversion of time dilation to distance are at $z = 0$. One needs to use equations (1), (2), and (5) to properly relate peak intensity to distance.

*Acknowledgments*: This work was done under the auspices of the US Department of Energy and was funded in part by the Gamma Ray Observatory Guest Investigator program. The BATSE data were prepared by J. Norris, R. Nemiroff, and J. Bonnell. We thank D. Band, J. T. Bonnell, R. Epstein, C. Ho, R. Nemiroff, J. Norris and N. Wright for helpful discussions and corrections. This paper benefited from many fruitful discussions at the Aspen Center for Physics.

# REFERENCES

Band, D. 1993, in Compton Gamma-Ray Observatory: St. Louis, 1992, eds. M. Friedlander, N. Gehrels, & D. J. Macomb (New York: AIP), 734.

Band, D. L, et al. 1993, Ap. J. 413, 281

Band, D. L. 1994 Ap. J. 432, L23

Band, D. L. 1994 et al. Ap. J. 434, 560

Davis, S. P., Norris, J. P., Kouveliotou, C., Fishman, G. J., Meegan, C. A., Pasiesas, W. C. 1994, in Gamma-Ray Bursts: Huntsville, 20-22 October, 1993, eds. G. J. Fishman, J. J. Brainerd, & K. C. Hurley (New York: AIP), p. 182.

Dermer, C. D. 1992 Phys. Rev. Lett 68, 1799

Epstein, R. I., Fenimore, E. E., Leonard, P. J. T., & Link, B., 1993 Annals NY Acad. Sci., 688, 565

Fenimore, E. E., et al. 1988 Ap. J. 335, L71

Fenimore, E. E., et al. 1992 Nature 357, 140

Fenimore, E. E., et al. 1993 Nature 366, 40

Fenimore, E. E., in 'T Zand, J. J. M., Norris, J. P., Bonnell, J. T., Nemiroff, R. J., 1995a submitted to ApJ Lett

Fenimore, E. E., et al. 1995b Ap. J., in preparation

Fishman, G., et al. 1992, in Gamma-Ray Bursts: Huntsville, 1991, eds. W. S. Paciesas & G. J. Fishman (New York: AIP) 13.

Kolb, E. W., & Turner, M. S. 1990 *The Early Universe* (Addison-Wesley, Redwood City, Calif.)

Link, B., Epstein, R. I., Priedhorsky, W. C. 1993 Ap. J. 408, L81

Loredo, T. and Wasserman, I. 1993, in Compton Gamma-Ray Observatory: St. Louis, 1992, eds. M. Friedlander, N. Gehrels, & D. J. Macomb (New York: AIP)751.

Mao, S. and Paczyński, B. 1992 Ap. J. 388, L45

Mészaros, P. and Rees, M. J. 1992 Ap. J. 397, 570

Meegan, C. A., et al. 1992 Nature, 355, 143

Meegan, C. A. et al. 1994 "The Second BATSE Burst Catalog." available from the Compton Gamma Ray Observatory Science Support Center.

Miller, G. S., Epstein, R. I., Nolta, J. P., Fenimore, E. E., 1991 Phys. Rev. Lett. 66, L395

Mitrofanov, I. G., et al. 1993, in Compton Gamma-Ray Observatory: St. Louis, 1992, eds. M. Friedlander, N. Gehrels, & D. J. Macomb (New York: AIP), p. 761.

Mitrofanov, I. G., 1994, in Gamma-Ray Bursts: Huntsville, 20-22 October, 1993, eds. G. J. Fishman, J. J. Brainerd, & K. C. Hurley (New York: AIP), p. 187.

Murakami, T. et al., 1988 Nature 335, 234

Nemiroff, R. J., et al. 1994 Ap. J. 435, L133

Norris, J. P., et al. 1994 Ap. J., 424, 540

Norris, J. P. et al. 1995 Ap. J., 439, 542

Norris, J. P. 1994, in Gamma-Ray Bursts: Huntsville, 20-22 October, 1993, eds. G. J. Fishman, J. J. Brainerd, & K. C. Hurley (New York: AIP), p. 177

Petrosian, V., Azzam, W. J. and Meegan, C. A. 1993, in Compton Gamma-Ray Observatory: St. Louis, 1992, eds. M. Friedlander, N. Gehrels, & D. J. Macomb (New York: AIP), 84

Piran, T. 1992 Ap. J. Lett. 389, L45

Podsiadlowski, P., Rees, M. J., and Ruderman, M. 1994, Cambridge University Astronomy preprint submitted to MNRAS.

Schaefer B. S. 1992, in Gamma-Ray Bursts Observations, Analyses, and Theories, ed. C. Ho, R. I. Epstein, & E. E. Fenimore (New York: Cambridge University Press), 107

Wickramasinghe, W. A. D. T., et al. 1993 Ap. J., 411, L55

Wijers, R. A. M. J., Paczyński, B. 1994 Ap. J. 432, L23

# FIGURE CAPTIONS

**Fig. 1:** a) Average $1+z$ and comoving distance versus standard candle luminosity $L_0$ for various brightness classes: the dotted, dashed, and dot-dash curves are for the bright, dim, and dimmest classes of Norris et al. 1994, respectively. The long dash curve is for typical bursts from PVO.

b) The ratio of the $1+z$ factors for three pairs of brightness classes: the dotted, dot-dash, and long dash curves are dimmest/bight, dim/bright, and bright/PVO, respectively. If there is no $W$-correction, Fig. 1b gives the relationship between observed time dilation and standard candle luminosity. The solid horizontal lines represent acceptable ranges ($\pm 1\sigma$) from various time dilation tests assuming no $W$-correction. The horizontal dotted line at $4.6 \times 10^{50}$ erg s$^{-1}$ represents the acceptable range of the standard candle luminosity consistent with the log $N$-log $P$ distribution.

c) Observed time dilation versus luminosity including the $W$-correction for the bandpass 25-115 keV. The solid horizontal lines represents the acceptable ranges of luminosity ($\pm 1\sigma$) from various time dilation tests if one includes the $W$-correction. If GRBs have the luminosity associated with the log $N$-log $P$ distribution, then one would expect to see only a factor of 1.3 time dilation between the dimmest and brightest BATSE GRBs rather than a factor of 2 (cf. panel c).

**Fig. 2:** The $W$-correction function, $W(yE_L, yE_U)/W(E_L, E_U)$, as a function of $y$ as determined by fitting a model of the time variation to the average pulse shapes. The solid, dotted, and dashed curves are for the 25 to 115 keV, 115 to 320 keV, 25 to 57 keV bandpass in BATSE for the average

pulse shape test. The long dashed curve is for the average auto-correlation test in the BATSE 115 to 320 keV bandpass. The points are values where the function can be estimated directly from the observations. The filled square is for the 25 to 57 keV bandpass at $y = 2$, the filled triangle is for the 115 to 320 keV bandpass at $y = 2.8$, and the open triangle is for the 115 to 320 keV bandpass at $y = 3$ for the average auto-correlation correction factor. These points demonstrate that our model is valid to at least $y \sim 3$.

**Fig. 3:** a) Time dilation versus luminosity for the 25 to 57 keV bandpass of BATSE. The curves are labelled as in Fig. 1c.
b) Time dilation versus luminosity for the BATSE bandpass 115-320 keV. The solid horizontal line is the acceptable range of luminosity ($\pm 1\sigma$) from the average auto-correlation test for time dilation assuming no $W$-correction. The dotted horizontal line at $4.6 \times 10^{50}$ erg s$^{-1}$ represents the acceptable range of the standard candle luminosity consistent with the log $N$-log $P$ distribution.
c) Time dilation versus luminosity suitable for the average auto-correlation test. The Dimmest-Brt and Dim-Brt curves use the $> 320$ keV energy range of BATSE. The PVO-Brt curve uses the $> 115$ keV energy range of BATSE and the 100 - 2000 keV energy range of PVO. The solid horizontal line is the acceptable range ($\pm 1\sigma$) from the average auto-correlation test for time dilation including the $W$-correction.

**Fig. 4:** The log $N$-log $P$ distributions for various standard candle luminosities compared to the BATSE and PVO observations assuming $q_0 = 1/2$. The solid curve ($L_0 = 4.6 \times 10^{50}$ erg s$^{-1}$) provides the best fit to the log

$N$-log $P$ observations. Only data above the threshold for the instruments were used: 1.0 photons s$^{-1}$ cm$^{-2}$ for BATSE and 20 photons s$^{-1}$ cm$^{-2}$ for PVO. The dotted curve ($L_0 = 2 \times 10^{51}$ erg s$^{-1}$) corresponds to the luminosity expected for an observed time dilation of 2.0 without an $W$-correction. The dashed curve ($L_0 = 10^{52}$ erg s$^{-1}$) corresponds to the luminosity expected for an observed time dilation of 2.0 with an $W$-correction. Although strong evolution might be able to force the distributions with large luminosities into agreement with the observations, it implies either a higher density of GRB progenitors and/or more luminous events in the distant past and that the observed $-3/2$ power law in PVO is a coincidence.

**Fig. 5:** a) The density and/or luminosity evolution necessary to achieve agreement with the BATSE-PVO log $N$-log $P$ distribution as a function of the standard candle luminosity. The density evolution is assumed to have the form $(1 + z)^{p_\rho}$ and the luminosity evolution has the form $(1 + z)^{p_L}$. The extent of the curve represents the range over which acceptable fits are possible. The solid line is for cosmologies with $q_0 = 1/2$ and the dotted line corresponds to $q_0 = 1$.

b) The $\chi^2$ for fitting the -3/2 power law portion of the BATSE-PVO log $N$-log $P$ distribution to cosmological models with different standard candle luminosities. For luminosities greater than $\sim 2 \times 10^{51}$ erg s$^{-1}$, it should be considered a coincident that the log $N$-log $P$ distribution shows a -3/2 power law since it arises because of the evolution, not because of a homogeneous distribution of sources.

**Fig. 6:** The average $1 + z$, $y$, and time stretching as a function of the standard candle luminosity, $L_0$, if $q_0 = 0$. The labelling is the same as in Figure 1. When $q_0 = 0$, the bright BATSE events have values of $z$ that are closer to the those of the dimmest events (see panel a). Thus, the time stretching without $W$-correction (panel b) is smaller than if $q_0 = 1/2$. The time stretching with $W$-correction (panel c) is smaller yet such that no luminosity within the range of our assumptions can give a time stretching as large as observed (i.e, 2.0). The horizontal dotted line at $4.6 \times 10^{50}$ erg s$^{-1}$ represents the acceptable range of the standard candle luminosity consistent with the log $N$-log $P$ distribution.

**Fig. 7:** The log $N$-log $P$ distributions for various standard candle luminosities compared to the BATSE and PVO observations assuming $q_0 = 1$. The solid curve ($L_0 = 3.7 \times 10^{50}$ erg s$^{-1}$) provides the best fit to the log $N$-log $P$ observations. The dotted curve ($L_0 = 7 \times 10^{51}$ erg s$^{-1}$) corresponds to the luminosity expected for an observed time dilation of 2.0 without an $W$-correction. The dashed curve ($L_0 = 2.75 \times 10^{51}$ erg s$^{-1}$) corresponds to the luminosity expected for an observed time dilation of 2.0 with an $W$-correction.

TABLE 1

Distances to Cosmological GRBs from Time Dilation Tests

| | | No $W(E_L, E_U)$ | | | with $W(E_L, E_U)$ | | |
|---|---|---|---|---|---|---|---|
| Test | $S$ | $L_{50}$ | $z_{\text{dimmest}}$ | $y$ | $L_{50}$ | $z_{\text{dimmest}}$ | $d$ (Gpc/$h_{75}$) |
| Peak Align[1] | $\sim 2.25$ | 22 | 2.5 | 3.6 | 126 | 9.5 | 5.5 |
| Fluence/Peak[1] | $2.0^{+0.50}_{-0.25}$ | $13^{+22}_{-6}$ | $1.8^{+1.5}_{-0.6}$ | 3.0 | $69^{+93}_{-38}$ | $5.6^{+8.2}_{-2.4}$ | $4.9^{+1.0}_{-0.8}$ |
| Wavelet[1] | $\sim 2.25$ | 22 | 2.5 | | | | |
| Pulse Width[2] | $1.8^{+0.65}_{-0.51}$ | $7.5^{+24}_{-6.7}$ | $1.4^{+1.8}_{-1.0}$ | 2.5 | $36^{+118}_{-33}$ | $3.5^{+9.4}_{-2.7}$ | $4.2^{+1.6}_{-2.2}$ |
| Cts in Peak[2] | $2.2^{+0.72}_{-0.44}$ | $20^{+44}_{-13}$ | $2.4^{+2.8}_{-1.1}$ | 3.5 | $113^{+133}_{-82}$ | $8.6^{+12.3}_{-5.4}$ | $5.4^{+0.9}_{-1.3}$ |
| Total Cts[2] | $2.0^{+0.62}_{-0.47}$ | $13^{+28}_{-10}$ | $1.8^{+2.0}_{-1.0}$ | | | | |
| Duration[3] | $2.2 \pm 0.2$ | $20^{+9}_{-7}$ | $2.4^{+0.6}_{-0.6}$ | | | | |
| Auto-Corr[4] | $2.0 \pm 0.2$ | $21^{+12}_{-9}$ | $2.5^{+0.8}_{-0.7}$ | 3.0 | $120^{+39}_{-54}$ | $9.0^{+4.4}_{-3.8}$ | $5.5^{+0.4}_{-0.7}$ |

[1]Norris et al. 1994

[2]Davis et al. 1994

[3]Norris et al. 1995

[4]Fenimore et al. 1995b

TABLE 2
Data for Fitting to the BATSE-PVO log $N$-log $P$

| $P_1$ | $P_2$ | Satellite | $\Delta N_{obs}$ |
|---|---|---|---|
| 1.00 | 1.26 | BATSE | 18 |
| 1.26 | 1.78 | BATSE | 20 |
| 1.78 | 2.82 | BATSE | 12 |
| 2.82 | 7.08 | BATSE | 26 |
| 7.08 | 31.6 | BATSE | 10 |
| 20.0 | 25.1 | PVO | 41 |
| 25.1 | 31.6 | PVO | 20 |
| 31.6 | 39.8 | PVO | 27 |
| 39.8 | 56.2 | PVO | 21 |
| 56.2 | 100 | PVO | 23 |
| 100 | 1000 | PVO | 14 |

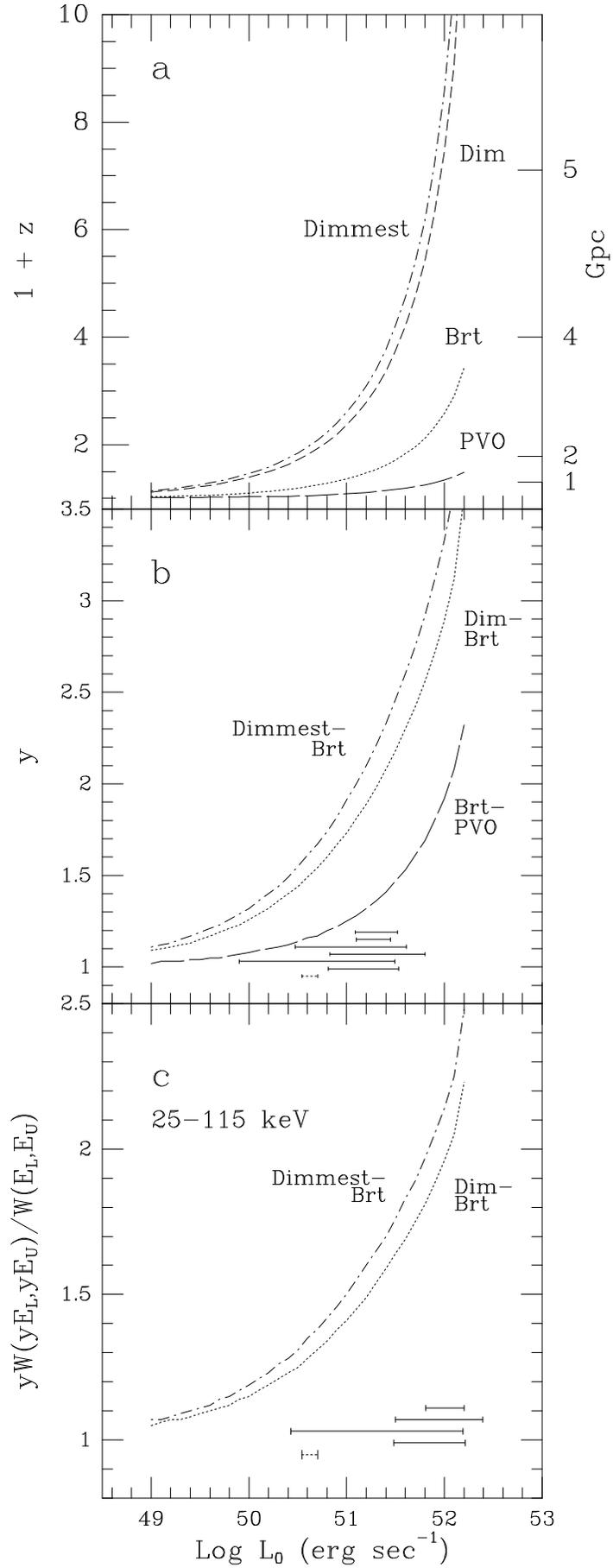

Fig 1

Fig 2

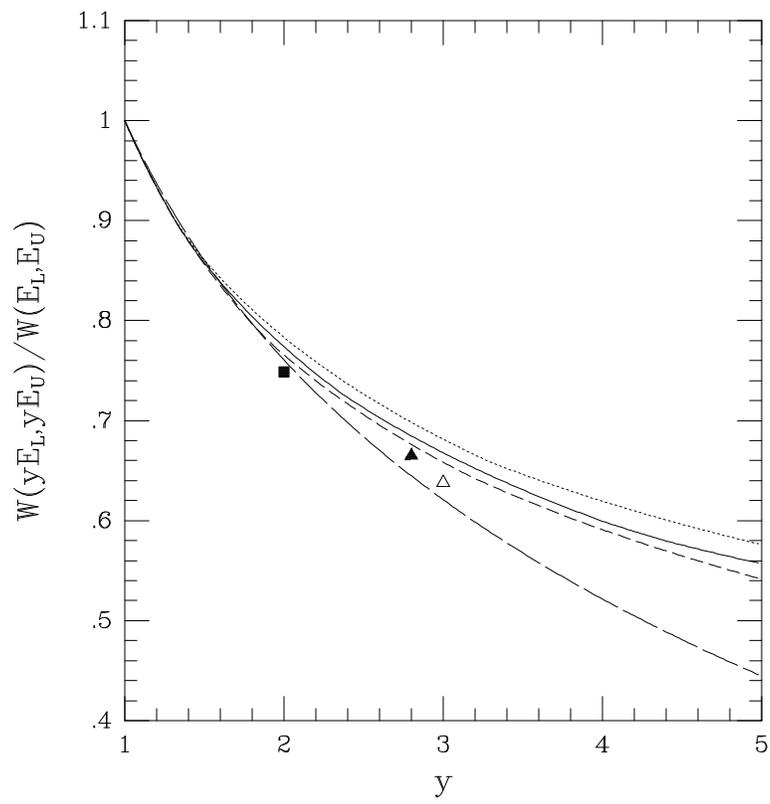

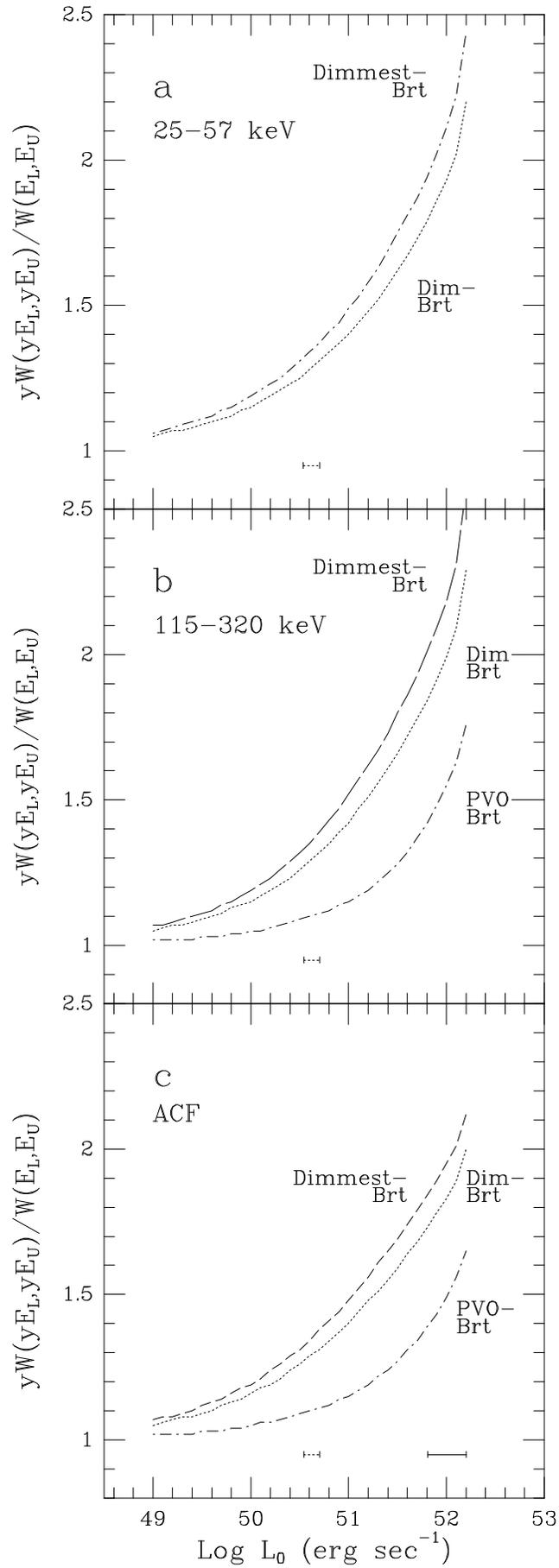

Fig 3

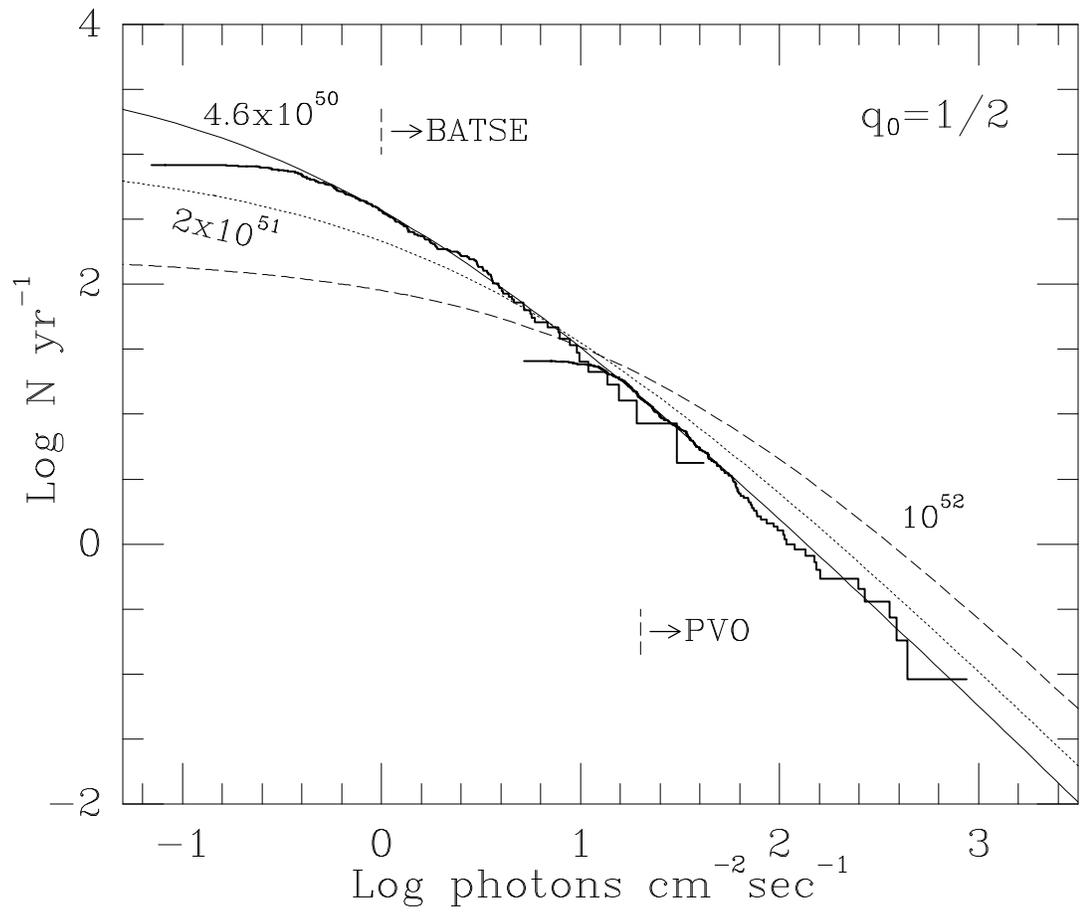

Fig 4

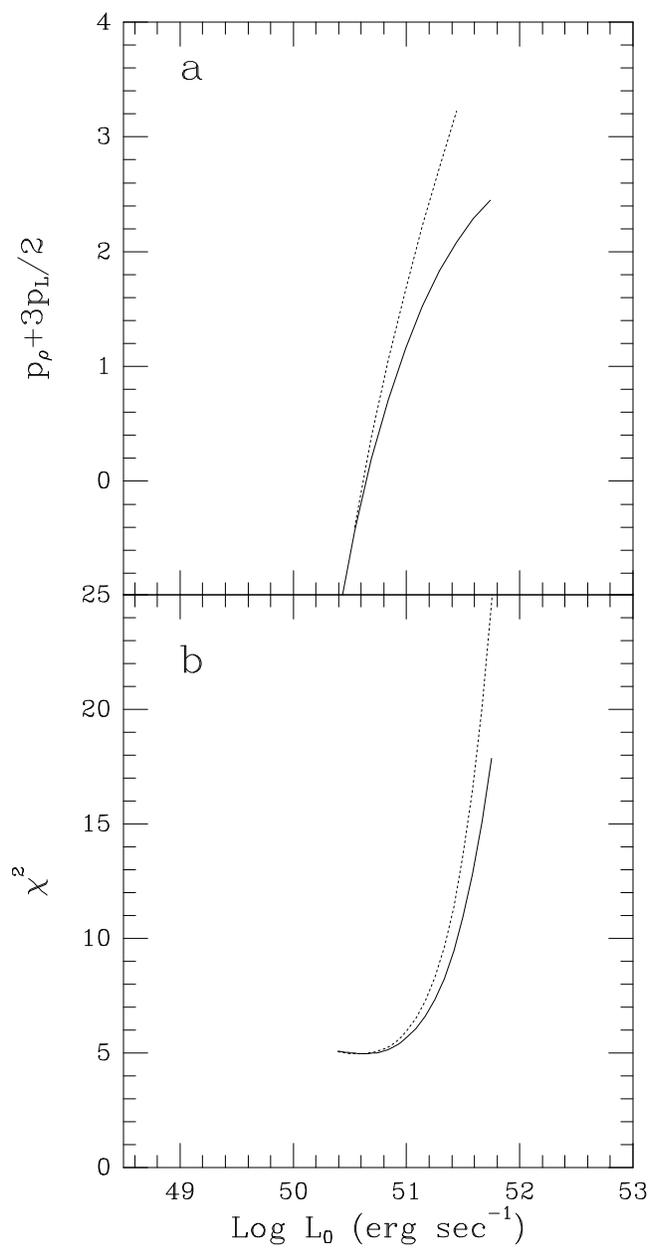

Fig 5

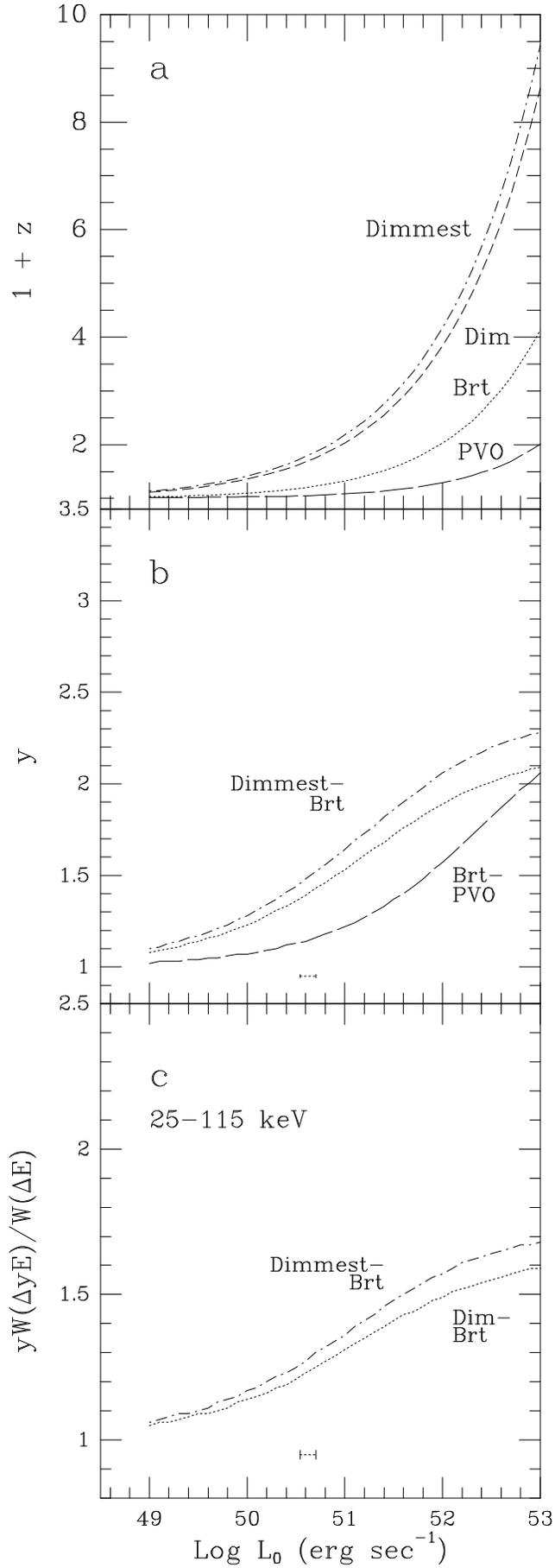

Fig 6

Fig 7

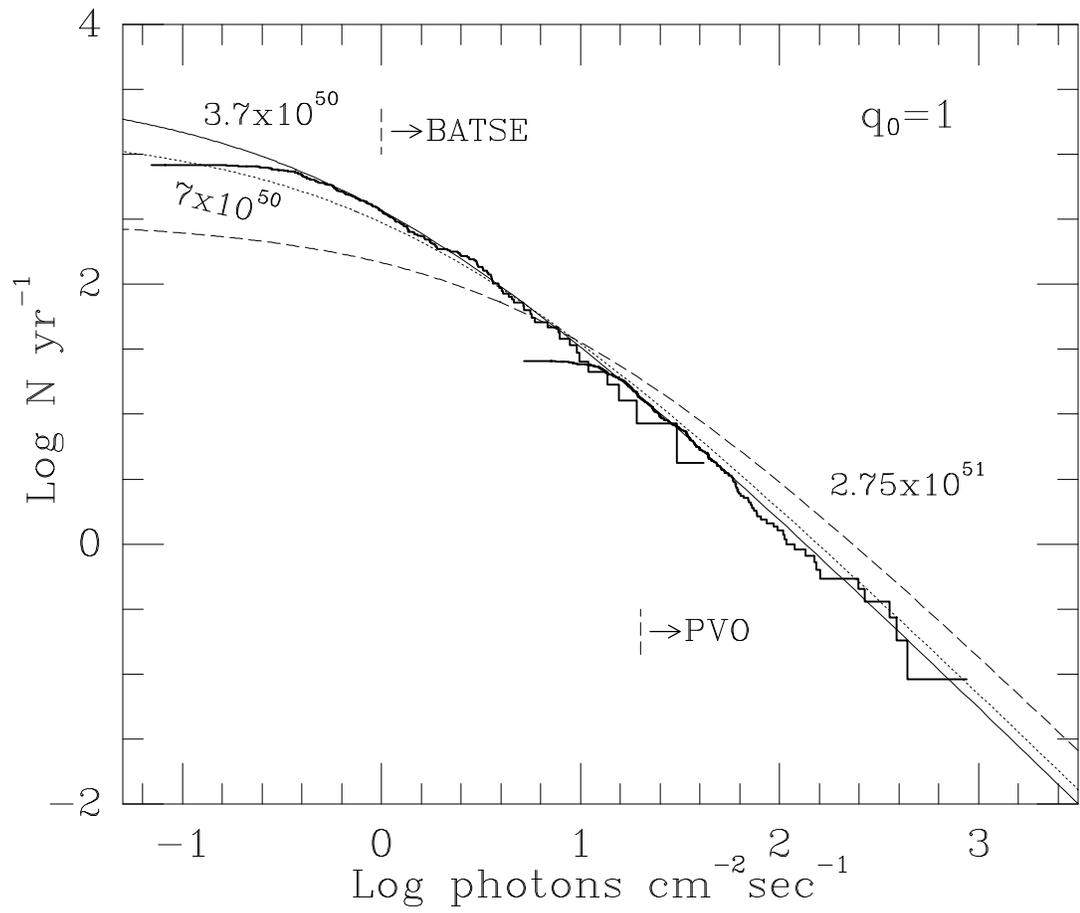

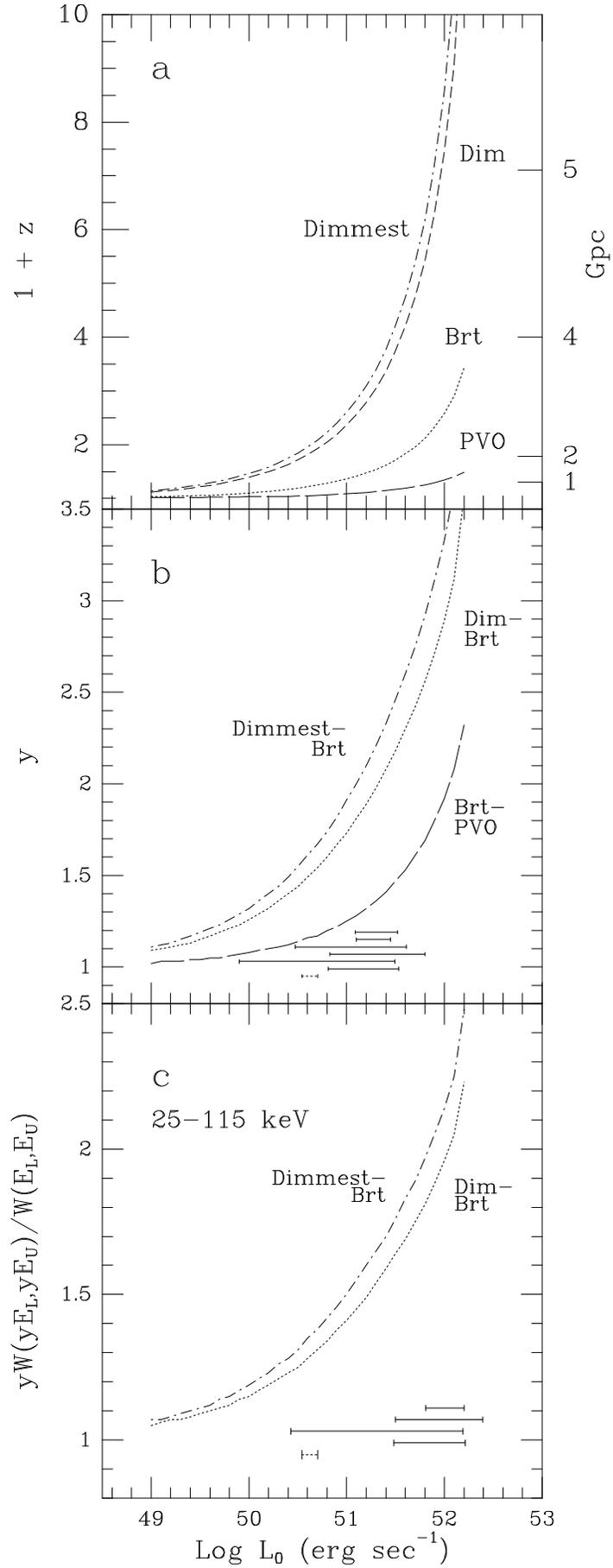

Fig 1

Fig 2

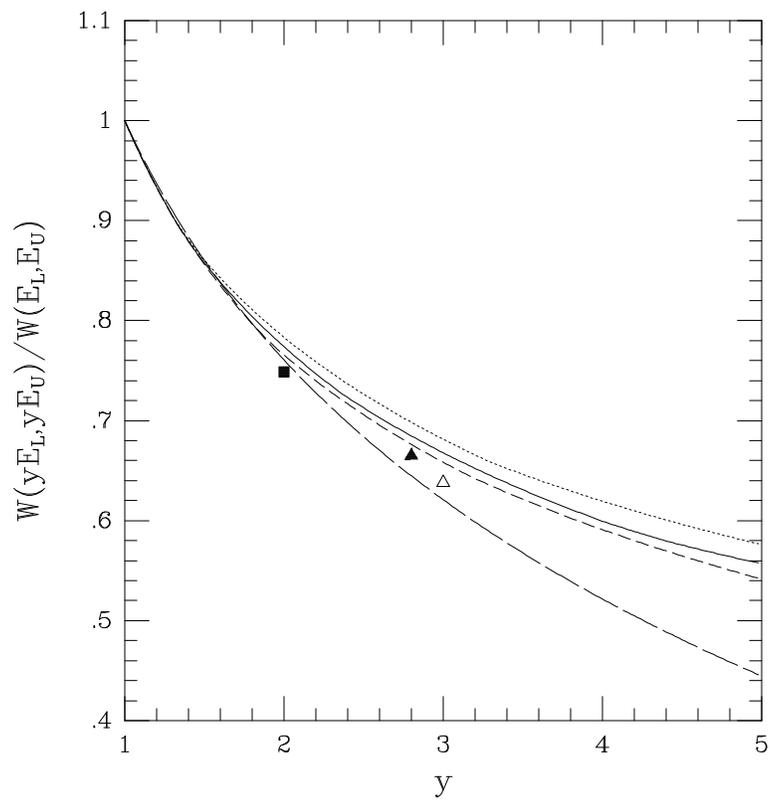

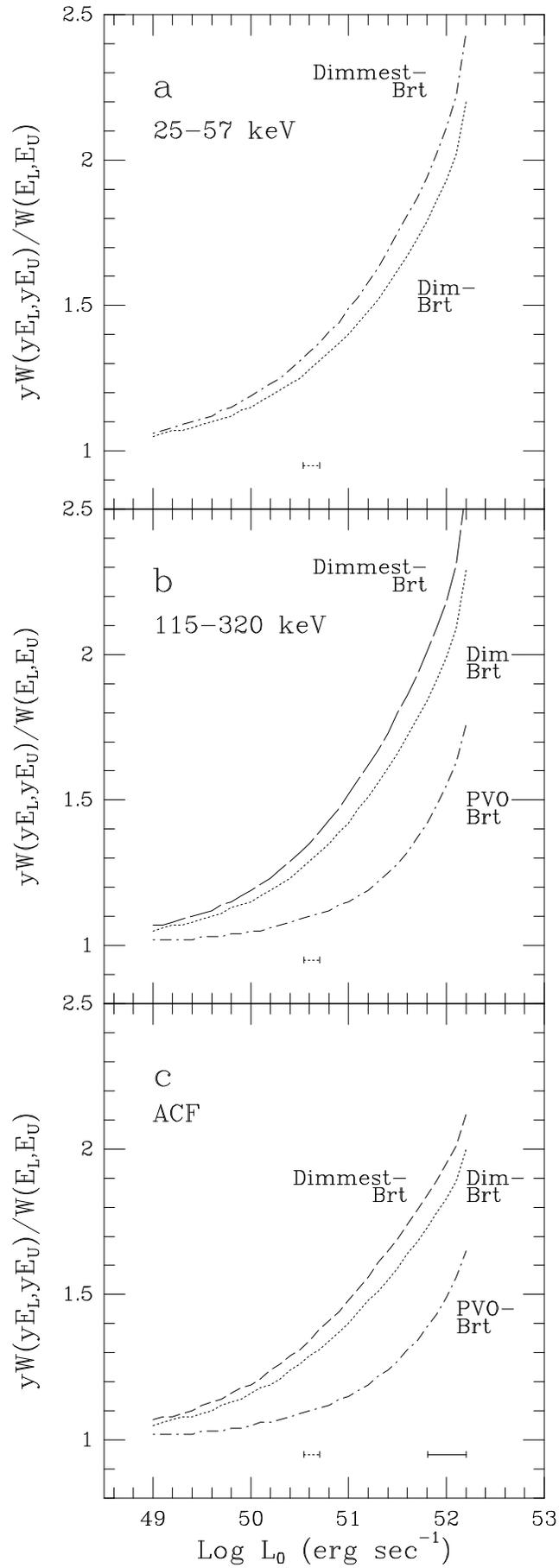

Fig 3

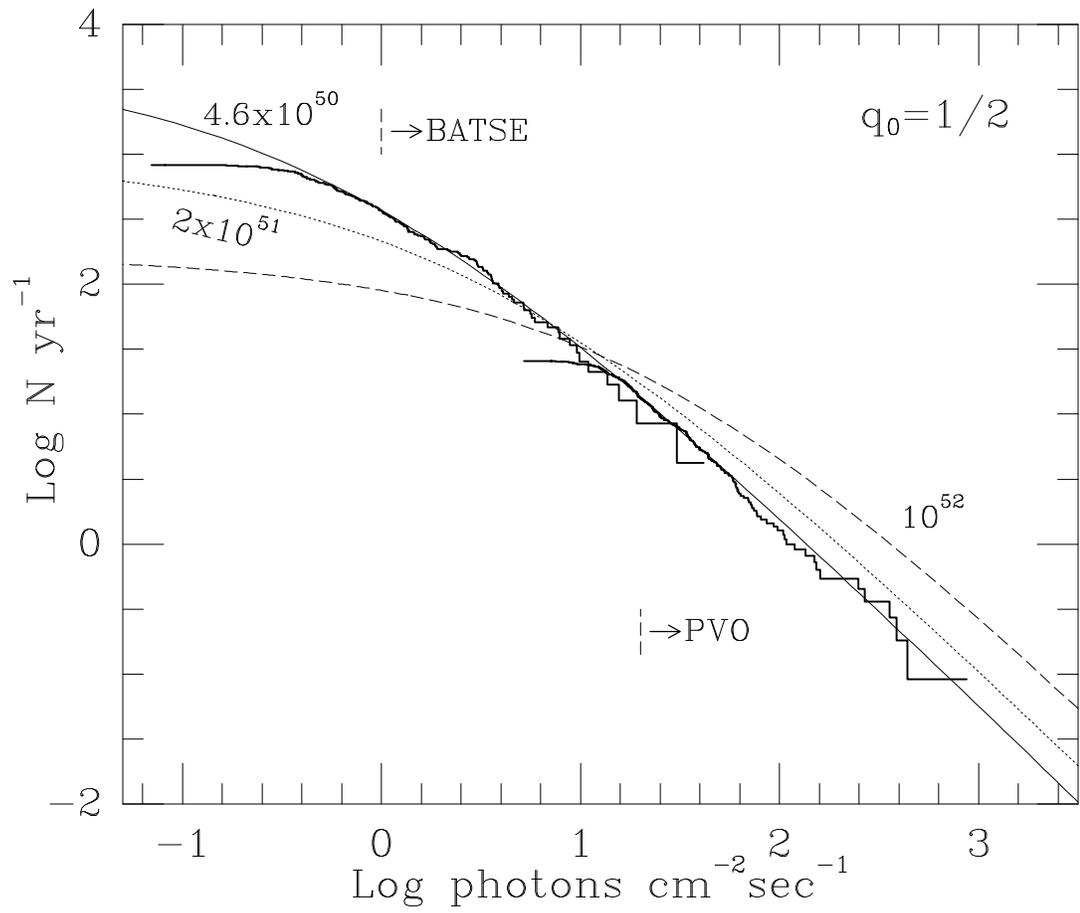

Fig 4

Fig 5

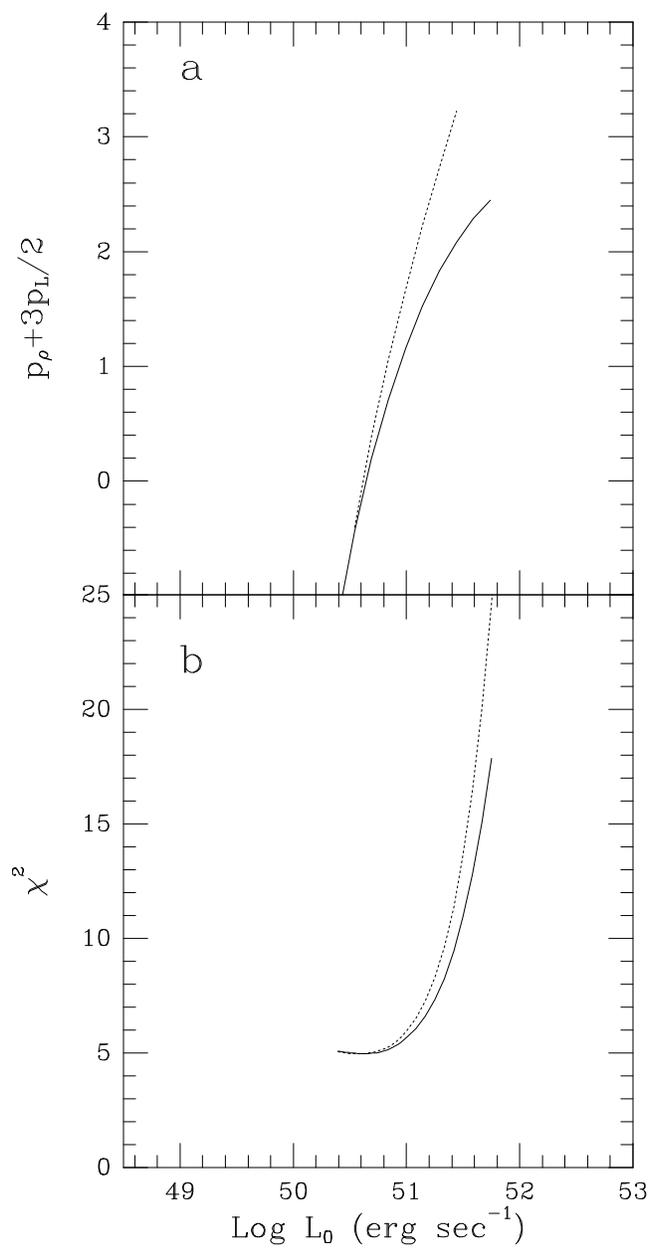

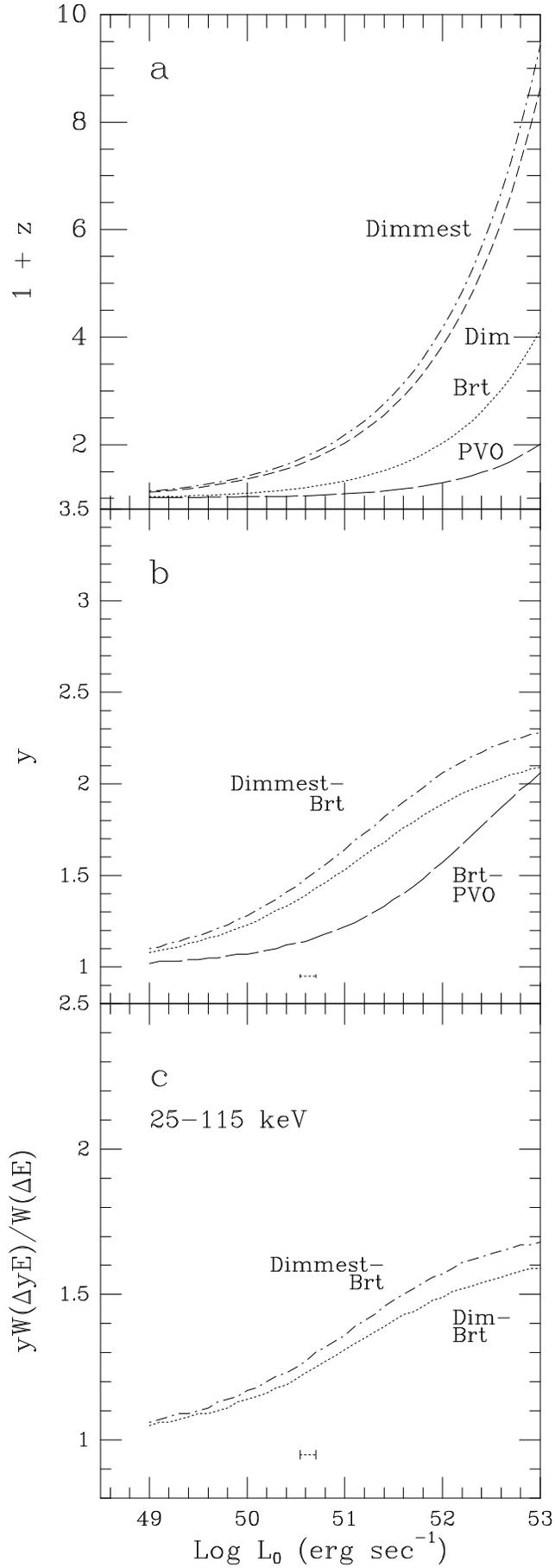

Fig 7

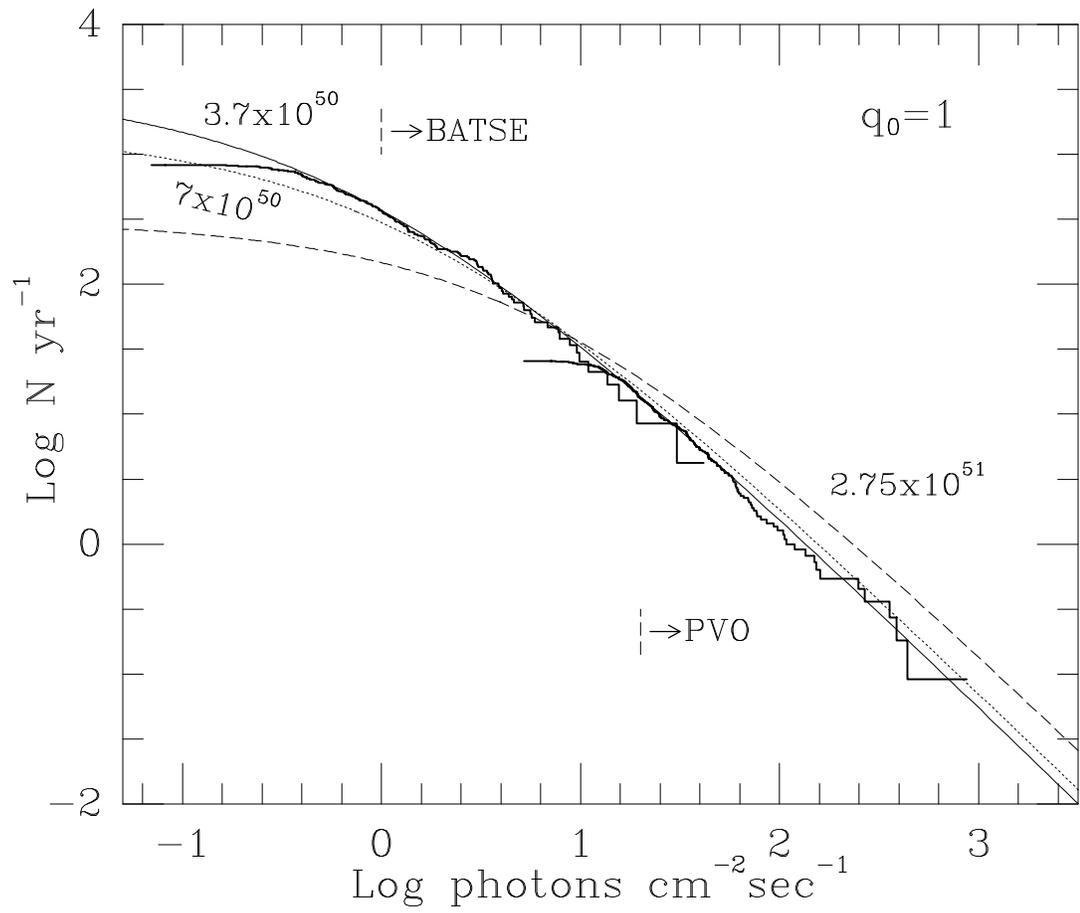